\documentclass[aps,reprint,prb,twocolumn,superscriptaddress,nofootinbib,longbibliography]{revtex4-2}

\usepackage{physics}
\usepackage[export]{adjustbox}
\usepackage{placeins}
\usepackage{fancybox,xcolor}
\def\SET{\text{\tiny SET}}
\begin{document}

%%%%%%%%%%%%%%%%%%%%%%%%%%%%%%%%%%%%%%%%%%%%%%%%%%%%%%%%%%

\title{Coupling--energy driven pumping through quantum dots: Role of coherences}

\author{Lukas Litzba}
\email{lukas.litzba@uni-due.de}
\affiliation{Fakult\"at f\"ur Physik and CENIDE, Universit\"at Duisburg--Essen, Lotharstra{\ss}e 1, Duisburg 47057, Germany}

\author{Gernot Schaller}
\affiliation{Helmholtz--Zentrum Dresden--Rossendorf, Bautzner Landstra{\ss}e 400, 01328 Dresden, Germany}

\author{J\"urgen K\"onig}
\affiliation{Fakult\"at f\"ur Physik and CENIDE, Universit\"at Duisburg--Essen, Lotharstra{\ss}e 1, Duisburg 47057, Germany}

\author{Nikodem Szpak}
\email{nikodem.szpak@uni-due.de}
\affiliation{Fakult\"at f\"ur Physik and CENIDE, Universit\"at Duisburg--Essen, Lotharstra{\ss}e 1, Duisburg 47057, Germany}

\date{\today}

\begin{abstract}
We study the impact of off--resonant tunneling and coherences on the electron pumping through quantum dots.
Thereby, we focus on two electron--pump setups where lowest--order tunneling processes are suppressed and the pump is exclusively driven by modulations of the coupling energy. 
The first setup is driven by switching on and off the couplings between the quantum dot and the leads, while the second setup employs measurements of the dot occupation.
We derive exact solutions for arbitrarily strong tunnel couplings in the absence of Coulomb interaction,
identify parameter regimes with optimal pumping currents or optimal energy efficiency,
and discuss similarities between both pumping mechanisms.
\end{abstract}

\maketitle

%%%%%%%%%%%%%%%%%%%%%

\section{Introduction}
\begin{figure}
\centering
    \includegraphics[width=\linewidth]{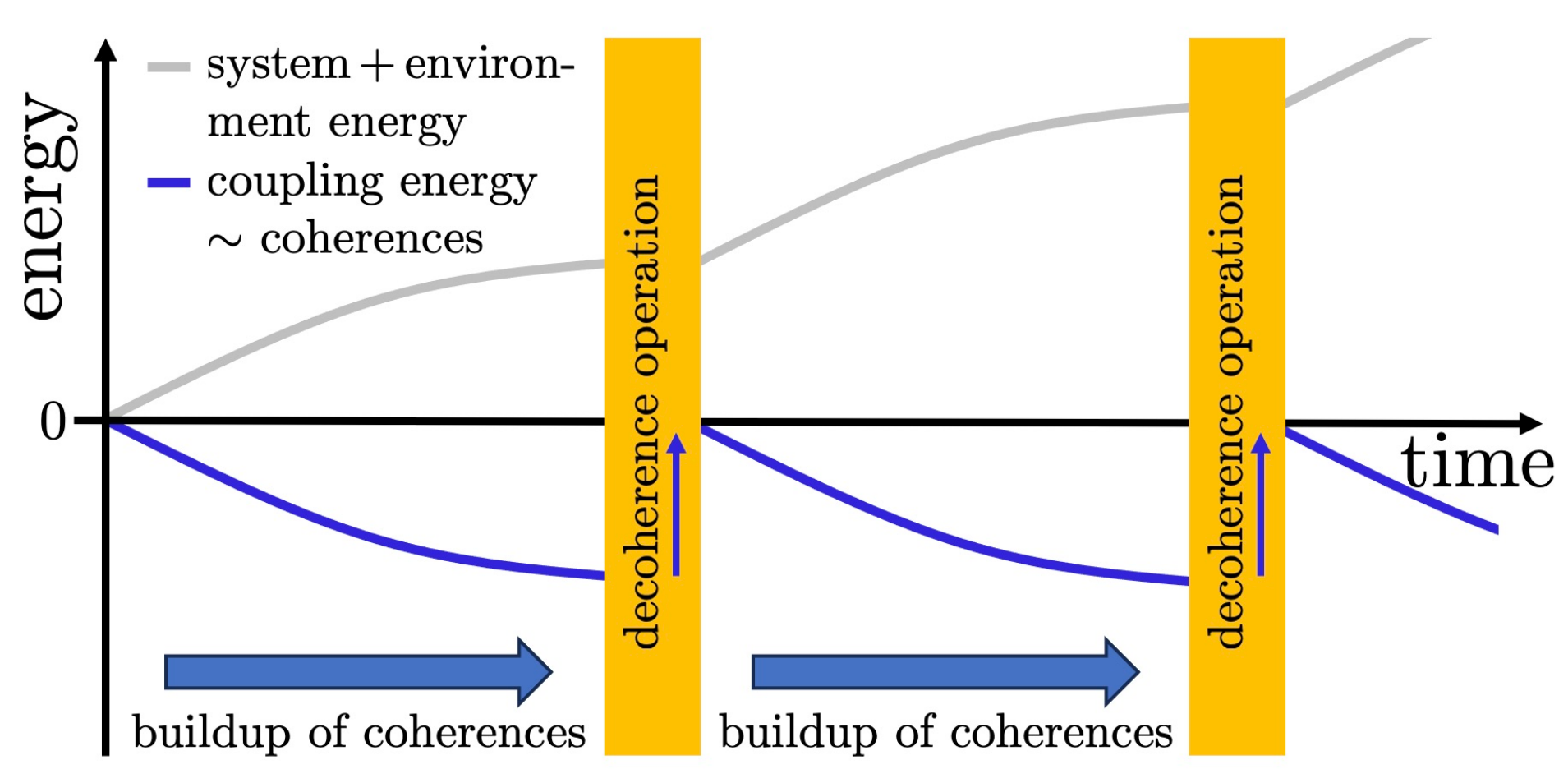}
    \caption{Sketch of an open quantum system subject to repeated interventions erasing the system--environment coherences and consequently the coupling energy (cf. Eq. \eqref{H_C}). In between, coherences may rebuild and the energy of system plus environment may increase on the cost of the coupling energy.
   }
   \label{fig:decoherenceoperation}
\end{figure}

\FloatBarrier
Modern miniaturized devices laying the foundations for quantum technologies are typically operating at the nanoscale where quantum effects begin to dominate.
A major challenge is to understand how quantum properties modify their operation and efficiency
\cite{EC_newman2017performance,wiedmannNonMarkovianDynamicsQuantum2020,EEmployingnon-Markovian2020,ENonMarkOtto2021, ESwitchingOttocycle2023,Improve_das2020quantum,Improve_mukherjee2020anti,Improve_xu2022minimal,ENon-Markovianthermal2022,picatosteDynamicallyEmergentQuantum2024,liu_periodically_2021,ECyclicQuantumEngines2023,katooraniFinitetimePerformanceCyclic2025,gattoQuantumOttoCycle2026,Improve_ghosh2019quantum,epump_RC,kalaee_Violating_2021,Improve_cavaliere2022dynamical,Improve_alamo2024minimal,EC_cangemi2024quantum}, compared to their classical counterparts.
From a theoretical point of view, such devices can be modeled as open quantum systems coupled to an environment. Often, the interaction between the quantum system and the environment is assumed to be weak and accounted for to lowest--order perturbation theory only.
That goes hand in hand with neglecting the coupling energy of the interaction. 
However, beyond the weak--coupling regime, the coupling energies and higher--order interaction processes become relevant \cite{EC_schulman2006ratcheting,EC_erez2008thermodynamic,EC_esposito2010entropy,EC_esposito2015nature,EC_strasberg2017quantum,EC_strasberg2019non,EC_trushechkin2022open,Teff,picatosteLocalGlobalApproaches2025,collaLocalEnergyAssignment2025,EC_perarnau2018strong}.
How they 
alter the properties and performance of the devices is currently a topic of intensive research
\cite{EC_perarnau2018strong,gelbwaser_klimovsky_work_2013,RC1,EC_newman2017performance,Feedback2018engelhardt2018maxwell,Feedback2018schaller2018electronic,elouardRevealingFuelQuantum2026,liuMaxwellsDemonQuantum2026,EC_bustos2019thermodynamics,wiedmannNonMarkovianDynamicsQuantum2020,EEmployingnon-Markovian2020,Improve_das2020quantum,liu_periodically_2021,ENonMarkOtto2021,Improve_xu2022minimal, ESwitchingOttocycle2023,ECyclicQuantumEngines2023,Improve_mukherjee2020anti,ENon-Markovianthermal2022,picatosteDynamicallyEmergentQuantum2024,katooraniFinitetimePerformanceCyclic2025,gattoQuantumOttoCycle2026}.
In contrast to previous works, where the impact of the coupling energy was treated on top of the lowest--order coupling processes, we focus here on devices that are enabled and operated exclusively by the coupling energy and the dynamical modulation of the associated coherences. 
They rely on the observation that \textit{the destruction of quantum coherence can pump energy into a system}
\cite{EC_schulman2006ratcheting}.
For tunneling interactions, in particular, the coupling energy is related to single--particle system--environment coherences. Consequently, repeated externally induced decoherence operations can continuously pump energy into the device by driving off--resonant tunneling (schematically shown in Figure \ref{fig:decoherenceoperation}).
Based on these repeated external operations leading to decoherence%
\footnote{In contrast to the concept of \textit{coherences as a resource} \cite{korzekwaExtractionWorkQuantum2016,chen_work_2025}, in which coherences are used as a fuel to extract energy from the bath, here we induce decays of the coherences between the bath and the QD. Such decays increase the coupling energy and finally the energy of the QD and the bath.},
we consider two pumping schemes of a device that pump electrons against an applied voltage:
For one of them, the coupling between the open system and the environment is periodically switched on and off
(cf. Section \ref{sec:scheme1}), while the other one is based on measurement--induced effects similar to the anti--Zeno effect \cite{AZ1,AZ2,AZ_fujii2010anti,QZuAZenoQD} (cf. Section \ref{sec:scheme2}).

The device we consider consists of a quantum dot~(QD), often described as an \textit{artificial atom}, placed between two fermionic baths. 
We consider the energy levels of the QD to be time--independent and higher than the chemical potentials of the baths. This ensures that lowest--order and resonant tunneling processes between the baths are blocked and only off--resonant tunneling processes related to the QD--bath coupling are relevant for the transport. 
In contrast, in common experimental realizations \cite{exp1999,exp2003,exp2007,exp2008,exp2012,exp_2014_1,exp2014_2,exp2014_3,exp2015_1,exp2015_2,exp2016,exp2017,exp2019_1,exp2019_2,exp2021,exp2022,exp2024}%
\footnote{In some experiments, only the gate voltage corresponding to the coupling strength is varied. In these cases, however, this results in an additional shift of the onsite potential of the QD and thus in a elevator--like behavior.},
pumping is achieved by varying the energy levels of the QD over time, as in an elevator, and involves primarily lowest--order and resonant tunneling processes.

In general, electron pumps have also been considered theoretically \cite{jauho1994time,Co_DeCo_JK_splettstoesser2006adiabatic,pump_croy2012nonadiabatic,Co_DeCo_JK_riwar2013zero,gurvitz_single_electron_2015,Ratchet2015,aginst_T_hussein2016heat,pump_croy2016full,Feedback2018schaller2018electronic,drag_lim2018engineering,pump_potanina2019optimization,epump_RC,Gurvitz_2019,Feedback_annby2024maxwell,sanchez_Making_2026,singh_Collective_2026}, have
important applications as single--electron sources in the field of metrology \cite{exp2019_1}, and are often treated in the weak--coupling limit. However, for strong couplings or for structured baths, higher--order or off--resonant tunneling processes and non--Markovian effects become also important \cite{epump_RC}. 
In this context, our focus on processes driven by the coupling energy itself is of particular interest because the work required to decouple QD and bath, or more generally, the work required to decouple the environment and the working medium of a thermodynamic machine, is a key cause of the reduction in efficiency of these devices at the nanoscale \cite{EC_newman2017performance,wiedmannNonMarkovianDynamicsQuantum2020,EEmployingnon-Markovian2020,ENonMarkOtto2021, ESwitchingOttocycle2023} --- in particular, in the strong--coupling regime.
Furthermore, performing local measurements on the device is influenced by the coupling energy which must be taken into account when studying measurement--induced pumping with feedback control, e.g. quantum Maxwell demons \cite{Feedback2018engelhardt2018maxwell,Feedback2018schaller2018electronic,elouardRevealingFuelQuantum2026,liuMaxwellsDemonQuantum2026}.
In addition, it is possible to drive an engine only by the work performed by the measurement process itself \cite{jussiauManybodyQuantumVacuum2023,sanchez_Making_2026,singh_Collective_2026}. In our work (second pumping scheme), we focus on measurement--induced pumping without feedback control by considering highly--structured baths.

For both pumping schemes, the structures of the baths and related non--Markovian effects \cite{zedlerWeakcouplingApproximationsNonMarkovian2009a} have significant influence on the efficiency. In particular, we find that non--Markovian effects can boost the performance of the electron pump which is already known for other engines \cite{Improve_mukherjee2020anti,EEmployingnon-Markovian2020,Improve_das2020quantum,wiedmannNonMarkovianDynamicsQuantum2020,ENonMarkOtto2021,Improve_cavaliere2022dynamical,Improve_xu2022minimal,ENon-Markovianthermal2022,ESwitchingOttocycle2023,picatosteDynamicallyEmergentQuantum2024}.
In contrast to the adiabatic regime, where the transferred charge per cycle is independent of the driving frequency for the coupling-- and decoupling--induced pumping (first pumping scheme) \cite{Co_DeCo_JK_splettstoesser2006adiabatic,gurvitz_single_electron_2015}, we operate in a deeply non--adiabatic regime.

We consider the QD in the absence of Coulomb interaction.
This allows us
to handle arbitrarily strong coupling and non--Markovian effects
because then an exact solution can be directly obtained by the extension of the exact Heisenberg equation and Laplace--transform technique \cite{Topp_2015} to time--dependent Hamiltonians.

The paper is divided mainly into two parts.
In the first part, we discuss pumping driven by decoherence operations that are extended in time and induced by a coupling-- and decoupling--procedure. In the second part, we study measurement--induced pumping which corresponds mainly to instant decoherence operations.
To be specific, we start with explaining the model 
used for both pumping schemes (cf. Section \ref{sec:model}). 
In Section \ref{sec:scheme1}, we consider the coupling-- and decoupling--induced pumping where in the following we will label it as pumping scheme 1. 
First, we discuss the slow pumping regime in Section \ref{sec:slowpumping}.
In particular, we identify a high energy efficiency regime and discuss efficiency improvements for finite cycle durations in a second step (cf. Section \ref{sec: improvefficency}). 
In Section \ref{sec:scheme2}, we consider measurement--induced pumping, where in the following we will label it as pumping scheme 2.
We study periodically repeated projective measurements (cf. Section \ref{sec:projecMea}), but also weak continuous measurement via a single--electron transistor (SET) (cf. Section \ref{sec:SET}).
In addition, we compare the results for pumping scheme 1 with those of pumping scheme 2. We demonstrate that both pumps are based on destruction of QD--bath coherences and show quantitative and qualitative similarities (cf. Section \ref{sec:scheme2vsCoUDeCou}).

%%%%%%%%%%%%%%%%%%%%%%%%%%%%%%%%%%%%%%%%%%%%%%%%%
\section{Model}
\label{sec:model}

Throughout this paper, we consider a quantum dot~(QD) placed between two fermionic baths, as sketched in Figure \ref{fig:RCQDRC} (left).

\subsection{Hamiltonian}

The starting point for our calculations is the single-impurity tunneling Hamiltonian
\begin{equation}\label{Htot}
    H(t)=H_\mathrm{S} + H_\mathrm{C}(t) + H_\mathrm{B} \, ,
\end{equation}
in which $H_\mathrm{S}= \varepsilon\, c^\dagger c$ with creation and annihilation operators $c^\dagger$ and  $c$ describes the open quantum system consisting of a single QD--orbital of energy $\varepsilon$ that is tunnel coupled to two fermionic baths modeled by $H_\mathrm{B}=\sum_{\nu,k} \varepsilon_{\nu,k} a^\dagger_{\nu,k} a_{\nu,k}$, where $\nu=1,2$ is the bath index and $\varepsilon_{\nu,k}$ the dispersion relation for states labeled by $k$ in bath $\nu$.
We assume the baths to be infinitely large and initially given by the tensor product of the Gibbs states $\rho_{\text{B},\nu}(0)\sim \exp\left({-\beta\sum_k (\varepsilon_{\nu,k}-\mu_\nu)a^\dagger_{\nu,k}a_{\nu,k}}\right)$ with inverse temperature%
\footnote{We set both bath temperatures equal for simplicity, however, they can be generally also treated as different.}
$\beta$
and with chemical potentials $\mu_\nu$. 
Without loss of generality, we consider $\mu_1\leq \mu_2$, define the voltage $V=\mu_2-\mu_1$ and choose $\mu_1=0$ as the energy reference.

In the following, we focus on the low--temperature limit, $\beta \to \infty$, and assume the energy of the QD orbital to lie above the chemical potentials of both leads, $\varepsilon >\mu_2$. 
Thereby, we ensure that resonant electron transport is suppressed.

The tunnel coupling between QD and bath is described by the Hamiltonian
\begin{equation}\label{H_C}
    H_\mathrm{C}(t) = \sum_{\nu,k}\left(\gamma_{\nu,k} (t)\,a_{\nu,k}^\dagger\,c + \text{h.c.}\right) \,.
\end{equation}
As such, it witnesses QD--bath coherences as the expectation value of $a_{\nu,k}^\dagger\,c$ corresponds to the coherences when representing the reduced density matrix for the QD and bath mode $(\nu,k)$ in the factorizing system-bath eigenbasis of $H_\text{S}+H_\text{B}$. Thus, in the following, we will refer to expectation values like $\left<a_{\nu,k}^\dagger\,c\right>$ as coherences.

In contrast to $H_\mathrm{S}$ and $H_\mathrm{B}$, we permit time--dependent $H_\mathrm{C}(t)$ through the tunnel matrix elements $\gamma_{\nu,k}(t)\equiv \gamma_{\nu,k} \, g_\nu(t)$, in which $g_\nu(t)$ accounts for the possibility of time--variable strength of the tunnel coupling between the QD orbital and bath $\nu$.
We assume that the associated spectral coupling--density
\begin{equation}
\begin{split}
     \Gamma_\nu (\omega, t)&=2 \pi \sum_k |\gamma_{\nu,k}(t)|^2\delta (\omega - \varepsilon_{\nu,k})\\
     &= g^2_\nu(t) \frac{\Gamma_\nu \Delta_\nu^2}{(\omega-\omega_\nu)^2+\Delta_\nu^2} 
\end{split}
\label{Gamma_nu(w,t)}
\end{equation}
has a Lorentzian shape with center position $\omega_\nu$, width $\Delta_\nu$ and the strength $\Gamma_\nu$ (for $g_\nu(t) \equiv 1)$, i.e., we use these three parameters instead of the infinitely many tunnel matrix elements $\gamma_{\nu,k}$ to characterize the tunnel coupling to bath $\nu$.

\subsection{Reaction--coordinate mapping}\label{sec:RC}

\begin{figure}[ht]
    \centering
    \includegraphics[width=\linewidth]{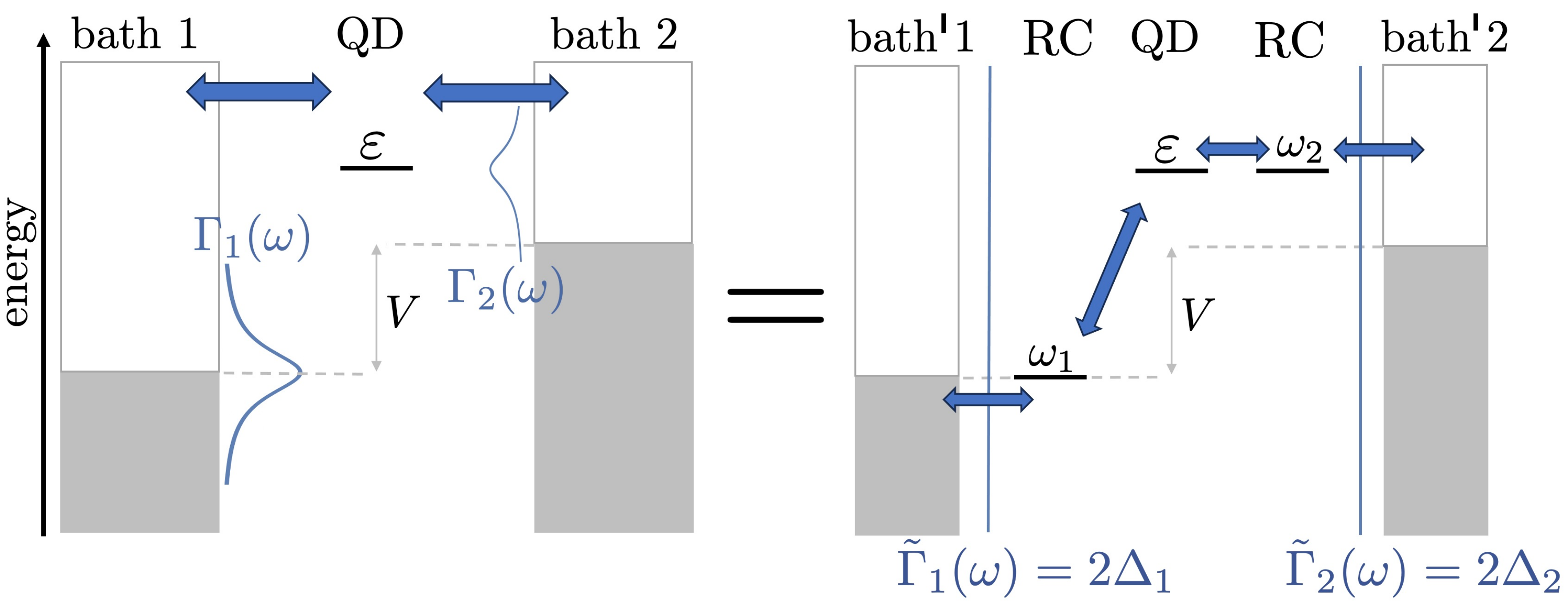}
    \caption{Schematic visualization of the fermionic reaction--coordinate mapping \cite{RC}: A quantum dot (QD) coupled to two Lorentzian shaped baths is equivalent to the situation where the QD is coupled to two reaction--coordinates (RC) that are coupled to two new baths with wideband spectral--coupling densities.  }
    \label{fig:RCQDRC}
\end{figure}

The total Hamiltonian $H$ (cf. Eq.~\eqref{Htot}) is only bilinear in the fermionic operators.
If, in addition, the Hamiltonian is also time independent then an exact solution is available by solving the Heisenberg equations of motion with the help of the Laplace transform \cite{Topp_2015}.
A direct application of this technique for our purpose is hindered by the time dependence of the coupling term $H_\mathrm{C}(t)$.
To circumvent this problem, we perform as a first step a fermionic reaction--coordinate mapping \cite{RC} for each of the baths, as visualized in Figure \ref{fig:RCQDRC}.
For each bath a unitary transformation is applied such that the QD orbital is no longer coupled to all the states of a given bath, but rather to one state only, which is referred to as the reaction coordinate (RC).
The coupling Hamiltonian $H_\mathrm{C}$ (cf. Eq.~\eqref{H_C}) together with a Lorentzian spectral coupling--density of width $\Delta_\nu$ and strength $\Gamma_\nu$ yields the creation operator 
\begin{equation}\label{rnud}
    r_\nu^\dagger=\sqrt{\frac{2}{\Delta_\nu \Gamma_\nu}}\sum_k \gamma_{\nu,k}a^\dagger_{\nu,k}
\end{equation}
for the reaction coordinate of bath $\nu$, where the prefactor ensures the normalization condition, and the reaction--coordinate energy is given by the center $\omega_\nu$ of the Lorentzian (cf. Eq. \eqref{Gamma_nu(w,t)}).
We stress that for factorized time--dependencies like in Eq. \eqref{Gamma_nu(w,t)}, one will always obtain a time--independent reaction--coordinate operator as the $g_\nu(t)$  cancels out (cf. Appendix A of Ref.~\cite{epump_RC}).
The remaining part of the bath Hilbert space is re--diagonalized, which leads to new bath--state creation operators $\tilde{a}^\dagger_{\nu,k}$ that couple to the reaction coordinate with new matrix elements $\tilde{\gamma}_{\nu,k}$.
As a result, the reaction--coordinate mapping rewrites the total Hamiltonian in the form
\begin{equation}
    H(t)=\widetilde{H}_\mathrm{S}(t) + \widetilde{H}_\mathrm{C} + \widetilde{H}_\mathrm{B} \, .
\end{equation}
The new super--system Hamiltonian
\begin{equation}
\label{tilde_H_S}
    \widetilde{H}_\mathrm{S}(t) = \varepsilon c^\dagger c + \sum_\nu \left[ \omega_\nu r_\nu^\dagger r_\nu + g_\nu(t)\sqrt{\frac{ \Gamma_\nu \Delta_\nu}{2}}(r_\nu^\dagger\,c +\text{h.c.})\right]
\end{equation}
now includes not only the QD level but also the two reaction coordinates.
It becomes time--dependent through the variation of the coupling strengths $g_\nu(t)$.
The new coupling Hamiltonian $\widetilde{H}_\mathrm{C}$ is now independent of time.
The couplings between new super--system and new baths are once more accounted for by a spectral coupling--density.
As has been shown in Ref.~\cite{RC}, the Lorentzian shape of the $\Gamma_\nu(\omega,t)$ of the original model yields a constant, i.e., time-- and energy--independent, spectral coupling--density $\tilde{\Gamma}_\nu = 2 \Delta_\nu$.
Its strength is given by twice the width of the original Lorentzian.
The originally structured bath coupling is, thus, mapped onto a structureless one, i.e., it fulfills the wideband condition, see Figure \ref{fig:RCQDRC}.
The time independence of $\tilde{\Gamma}_\nu$ makes it possible to extend, in a straightforward way, the Laplace--transform technique of Ref.~\cite{Topp_2015} to the system of consideration in the present paper, see Appendix \ref{sec: nonhermitianOperatorAppendix}.

%%%%%%%%%%%%%%%%%%%%%%%%%%%%%%%%%%%%%%%%%%%%%%%%%%%%%%%%%%%%%%%%%%%%%%%%%%

\section{ Scheme 1: coupling/decoupling induced pumping}\label{sec:scheme1}
Both pumping schemes that we discuss in this paper rely, on the one hand, on the structured spectral coupling densities, characterized by the center positions $\omega_\nu$, widths $\Delta_\nu$ and the strengths $\Gamma_\nu$ for bath $\nu = 1,2$ and, on the other hand, on the alternation of formation and decay of coherences between the quantum dot (QD) and the baths.

\begin{figure}
\centering
    \includegraphics[width=\linewidth]{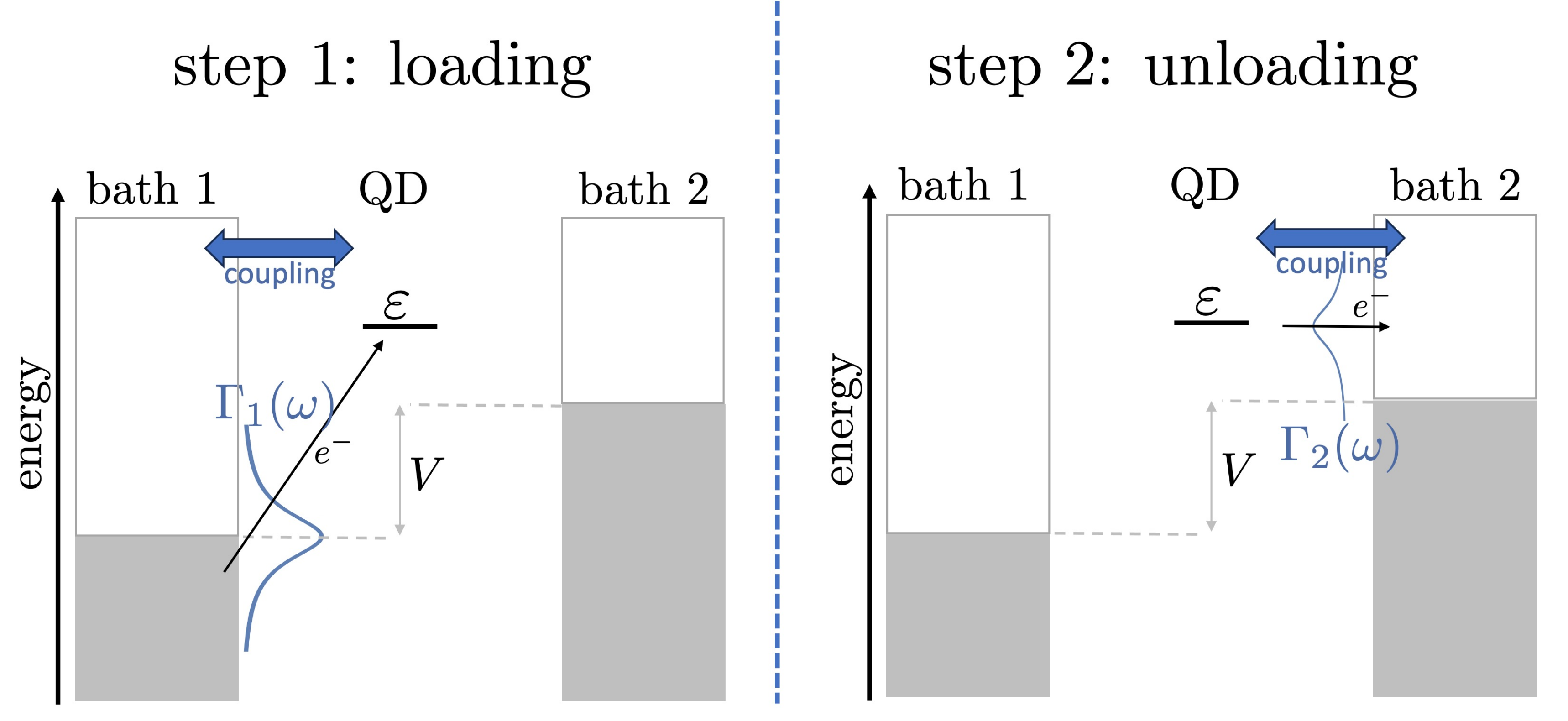}
    \caption{Schematic visualization of the functionality (step 1 and step 2 form one pumping cycle) of pumping scheme 1: The pump consists of a quantum dot (QD) that has a time--independent onsite potential $\varepsilon$ and is located between two baths at zero temperature. The difference of the chemical potentials $\mu_1$ and $\mu_2$ of the baths is given by $V$.  Here, the pumping results only from the coupling and decoupling processes with the baths. In order to study non--Markovian effects, we consider structured baths with Lorentzian shaped spectral densities $\Gamma_1(\omega)$, $\Gamma_2(\omega)$ \cite{zedlerWeakcouplingApproximationsNonMarkovian2009a}.}
    \label{fig:Schaubild2}
\end{figure}

\begin{figure}
    \centering
    \includegraphics[width=\linewidth]{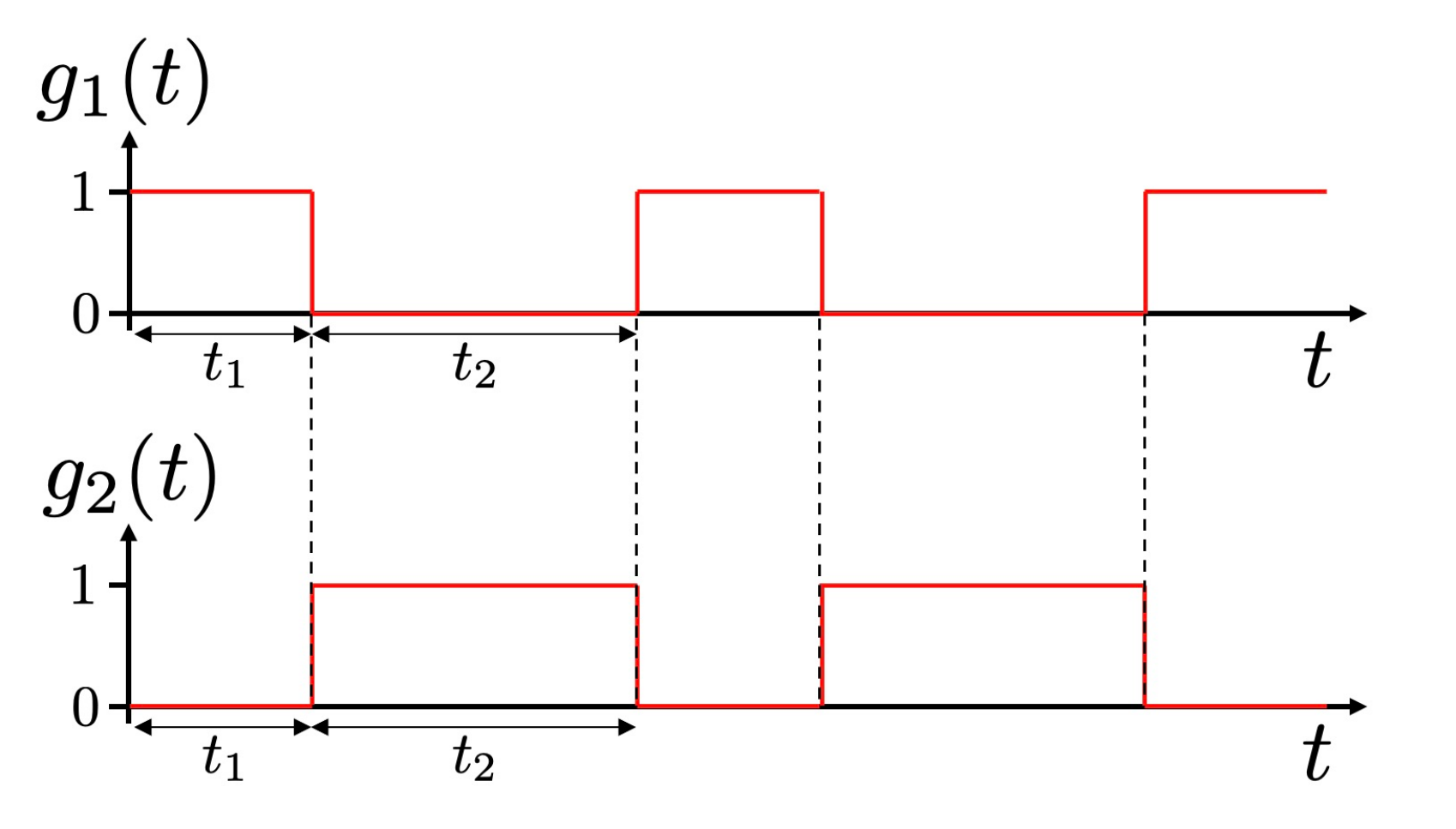}
    \caption{The coupling function $g_\nu (t)$ for $\nu=1,2$ as a function of time $t$ with the duration $t_1$ of step 1 and the duration $t_2$ of step 2. The functions fulfill the relation $g_1(t)=1-g_2(t)$.}
    \label{fig:g(t)}
\end{figure}

In the first pumping scheme, this alternation is triggered by periodically switching on the coupling to one of the baths and simultaneously switching off the coupling to the other bath (cf. Figure \ref{fig:Schaubild2}).
To be specific, we assume a stepwise%
\footnote{For computational reasons, we only consider sharp step functions for $g_\nu (t)$, but it is not necessary for the pump effect itself. 
Thus, we are always in a non--adiabatic regime.}
switching between the QD being coupled to bath 1 (with duration $t_1$) and to bath 2 (with duration $t_2)$.
This is described by the functions $g_\nu(t)$ as shown in Figure \ref{fig:g(t)}.

During step 1 and 2, charge can only be exchanged between the QD and bath 1 and 2, respectively.
The average number of electrons pumped into bath 2 is, therefore, the QD occupation at the beginning minus the occupation at the end of step 2.
This results in 
\begin{equation}
    N_{\text{pump}}=\left<n_\text{QD}(\ell t_1+(\ell-1)t_2)\right> -\left<n_\text{QD}(\ell t_1+\ell t_2)\right>
    \label{Npump}
\end{equation} 
for the $\ell$-th cycle, and the corresponding averaged current is given by
\begin{equation}
    I_\text{pump}= \frac{N_\text{pump}}{t_1+t_2} \, .
\end{equation}

%%%%%%%%%%%%%%%%%%%%%%%%%%%%%%%%%%%%%%%%%%%%%%%%%%%%%%%%%%%%%%%%%%%%%%%%%%
\subsection{Slow Pumping}
\label{sec:slowpumping}
In this section, we consider coupling periods $t_1$ and $t_2$ that are long enough for the device to reach the stationary state during each step (formally $t_1, \,t_2\rightarrow \infty$). 
Here, after each step, the coherences between the QD and the previously uncoupled bath have completely decayed. 
This corresponds to the (time--extended) decoherence operation from Figure \ref{fig:decoherenceoperation} that drives off-resonant tunneling.
In this case, the QD occupations entering Eq.~\eqref{Npump} are the stationary ones, $\left<n_{\text{stat},1}\right>$ and $\left<n_{\text{stat},2}\right>$, for the QD being coupled to bath 1 or 2, respectively.
The number of pumped charges per cycle is, then,
\begin{equation}\label{Npumstat}
    N_\text{pump} \approx \left<n_{\text{stat},1}\right>-\left<n_{\text{stat},2}\right> \, .
\end{equation}
In the weak--coupling limit for both baths, the stationary QD occupations are both zero, because we consider zero bath temperatures and $\varepsilon >\mu_2>\mu_1$. 
For stronger coupling, however,
$\left<n_{\text{stat},1}\right>$  and $\left<n_{\text{stat},2}\right>$ are finite, given by 
\begin{equation}\label{statnQD}
    \left<n_{\text{stat},\nu}\right>=  \int_{-\infty}^{\mu_\nu} \frac{\text{d}\omega}{2 \pi} \frac{ \Gamma_\nu }{(\varepsilon-\omega)^2+\left(\frac{\Gamma_\nu}{2}+\frac{(\varepsilon-\omega)(\omega-\omega_\nu)}{\Delta_\nu}\right)^2}
\end{equation}
(cf. Appendix \ref{sec: exactAppendixstationaryQDn}) where the coupling strength $\Gamma_\nu$ can be interpreted as an effective broadening of the energy level. Thus, $\left<n_{\text{stat},\nu}\right>$ increases for increasing coupling strength $\Gamma_\nu$ but decreases for increasing $\varepsilon-\mu_\nu$.
This implies that for equally coupled baths, $\Gamma_1(\omega)=\Gamma_2(\omega)$, since $\varepsilon - \mu_1 > \varepsilon - \mu_2$, there will be no transport against the bias, $N_\text{pump}<0$.
However, if the coupling strength $\Gamma_1$ is sufficiently large compared to $\Gamma_2$ then $\left<n_{\text{stat},1}\right> > \left<n_{\text{stat},2}\right>$ is possible, such that electrons are pumped against the applied voltage, $N_\text{pump}>0$.

As a further simplification, we consider an extremely weak coupling $\Gamma_2$ between the QD and the second bath, still respecting $\Gamma_2 t_2 \gg 1$, such that the QD is completely emptied after step 2.
The number of pumped electrons is, then, given by
\begin{equation}
    N_\text{pump}\approx\left<n_{\text{stat},1}\right> \, ,
\end{equation}
i.e., the stationary QD occupation of step 1.

\begin{figure}[ht]
    \centering
    \includegraphics[width=0.9\linewidth]{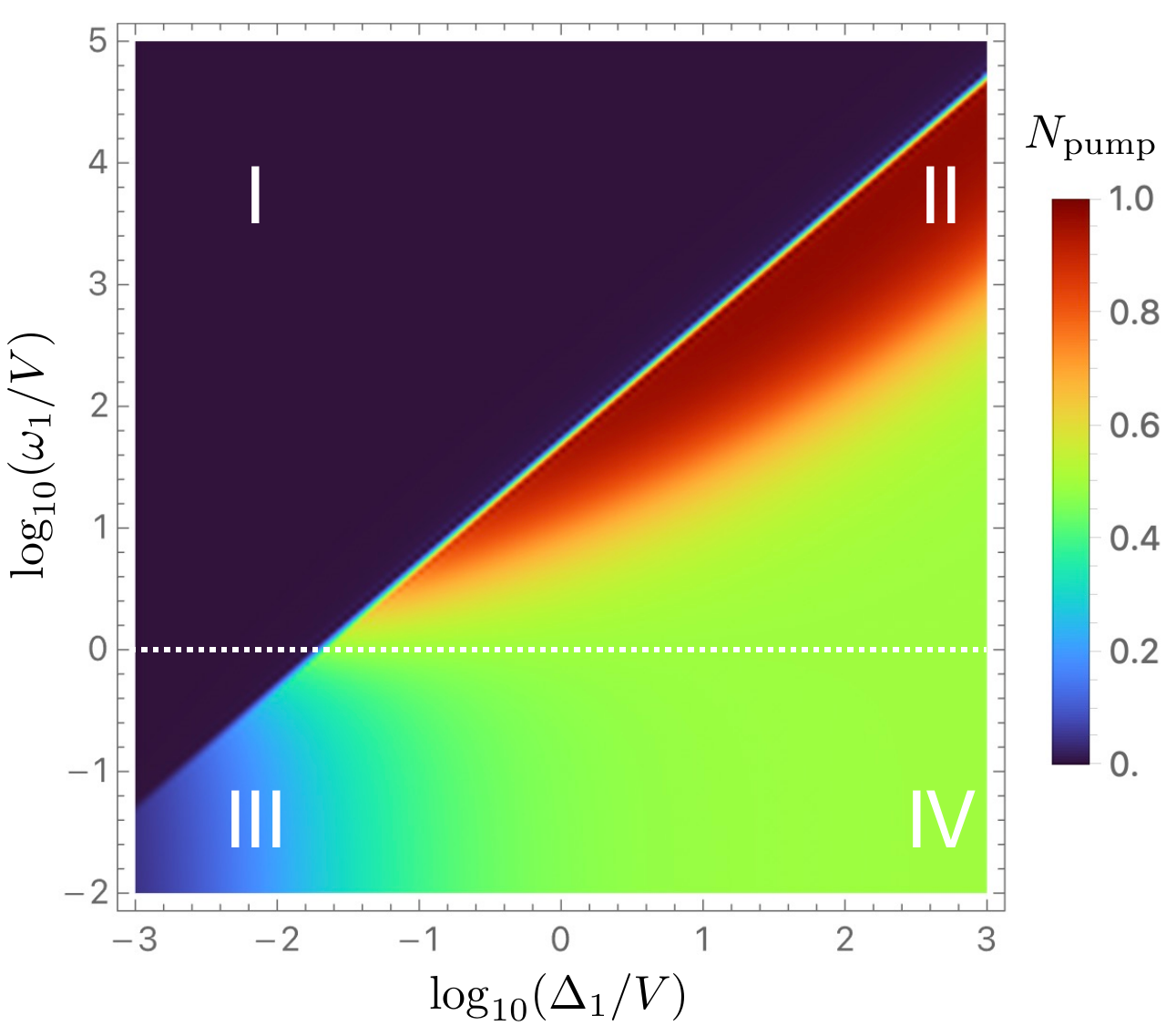}
    \caption{Average number of electrons $N_\text{pump}\approx\left<n_{\text{stat},1}\right> $ pumped during one cycle as a function of the width $\Delta_1$ (logarithmic scale) and the peak position $\omega_1$ (logarithmic scale) of the {spectral coupling--density} of the first bath in the slow--pumping regime, $t_1,t_2 \rightarrow \infty$. The QD--orbital energy $\varepsilon=V+0^+$ is directly above the chemical potential of the second bath. The coupling with the first bath is $\Gamma_1=100\,V$ whereas the coupling with the second bath is chosen  
    sufficiently weak (formally $\Gamma_2\searrow 0$) such that the QD is completely emptied after step 2.
    Pumping is suppressed for $\omega_1 > \frac{\Delta_1 \Gamma_1}{2 \varepsilon}$ (region I) and is largest in region II.
    }
    \label{fig:2D_N}
\end{figure}
\begin{figure}[ht]
    \centering
    \includegraphics[width=\linewidth]{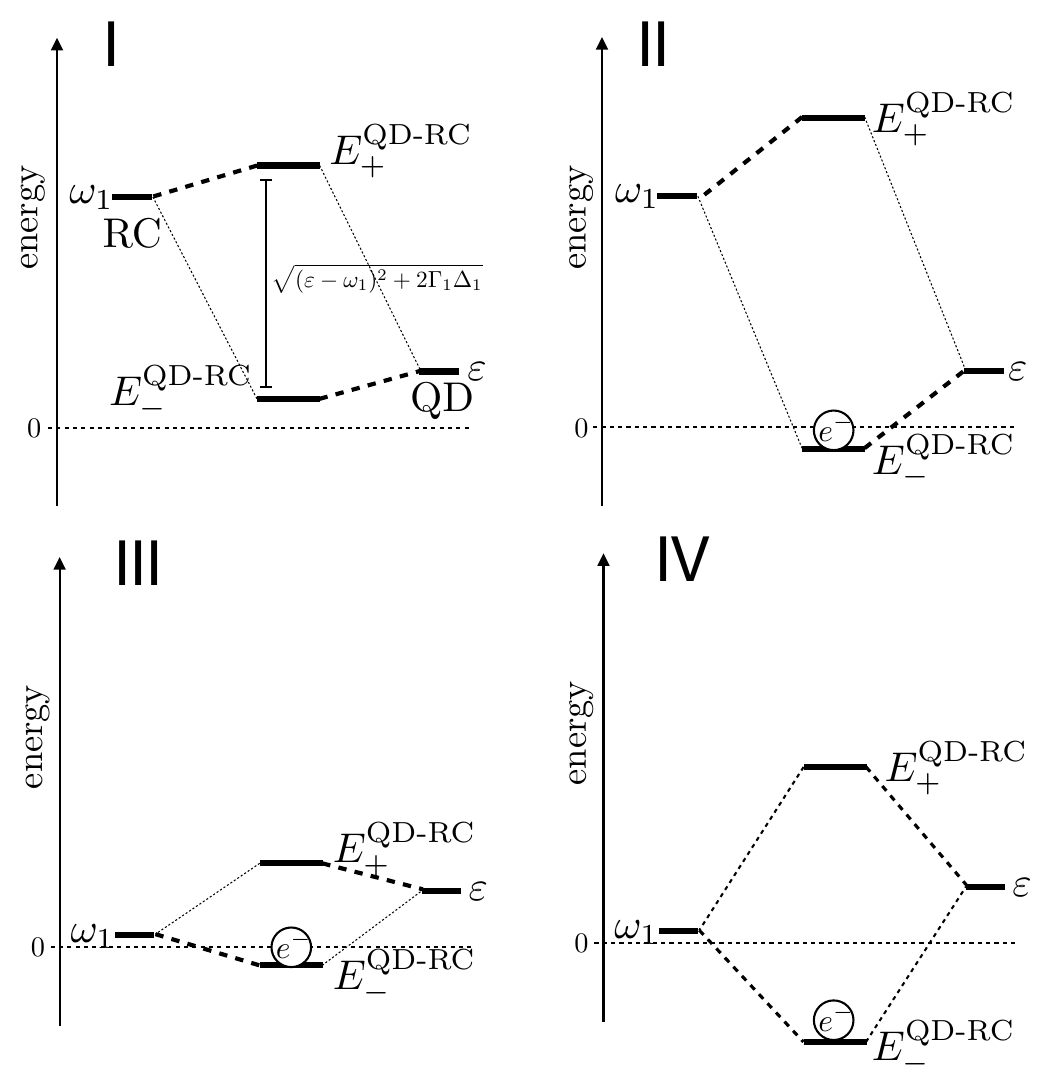}
    \caption{
    Schematic visualization of the hybridized states that result form coupling the QD and the reaction coordinate (RC): Each graph (I--IV) shows the energy levels of the reaction coordinate $\omega_1$ and of the QD $\varepsilon$ in relation to the chemical potential $\mu_1=0$. In addition, the two energy levels $E^\text{QD-RC}_\pm$ of the hybridized states that result from the coupling between the QD and the reaction coordinate of the first reservoir during step 1 with level splitting $\sqrt{(\varepsilon-\omega_1)^2+2 \Gamma \Delta_1}$ are drawn between the energy levels of the reaction coordinate and the QD. Furthermore, the thickness of the dashed lines that connect the pure QD/reaction coordinate and the hybridized levels symbolize the corresponding overlaps between them. The graphs I--IV correspond to the regions I--IV in Figure \ref{fig:2D_N}. 
    In region I ($\omega_1 >\frac{\Gamma_1 \Delta_1}{2 \varepsilon}$), the lowest hybridized energy is larger than $\mu_1$ and therefore 
    this state will not be significantly occupied.
    In contrast, in regions II--IV, the lowest hybridized energy levels lie below $\mu_1$ and these states will be predominantly occupied. 
    Since the overlaps between the lower hybridized states and the QD are different, so are the occupations of the QD:
    In region II, large overlap $\Rightarrow \left<n_{\text{stat},1}\right> \approx 1 $; in region III, small overlap $\Rightarrow $ $\left<n_{\text{stat},1}\right> \approx 0$; in region IV, approximately equal overlap with QD and reaction coordinate $\Rightarrow \left<n_{\text{stat},1}\right> \approx \frac{1}{2}$; dashed--white line in Figure \ref{fig:2D_N} ($\omega_1=\varepsilon$), exactly equal overlap with QD and reaction coordinate $\Rightarrow \left<n_{\text{stat},1}\right> =\frac{1}{2}$. 
    }
    \label{fig:RC_schematic_states}
\end{figure}
 \begin{figure}
     \centering
    \includegraphics[width=0.9\linewidth]{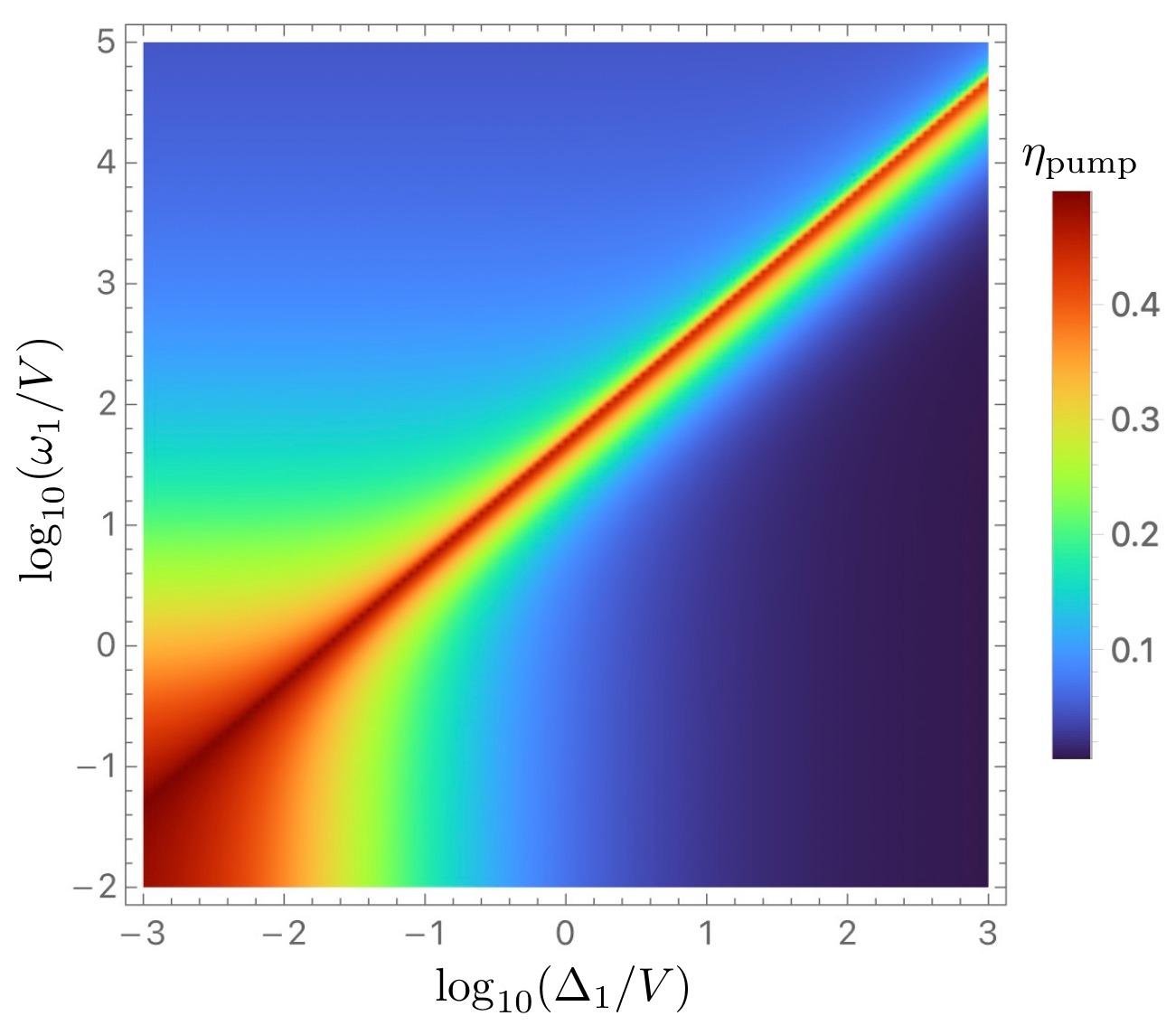}
    \caption{The energy efficiency $\eta_\text{pump}$ as a function of the width $\Delta_1$ (logarithmic scale) and the peak position $\omega_1$ (logarithmic scale) of the {spectral coupling--density} of the first bath. The parameters are the same as in Figure \ref{fig:2D_N}.
    In contrast to $N_\text{pump}$ that vanishes for highly structured baths ($\Delta_1 \ll V$), the efficiency $\eta_\text{pump}$ is optimized in this regime for $\omega_1 \ll V$.
    For $\omega_1= \frac{\Delta_1 \Gamma_1}{2 \varepsilon}$, $\eta_\text{pump}$ shows a special behavior analogous to $N_\text{pump}$ and reaches large values compared to $\eta_\text{pump}$ for the surrounding parameters.
    Moreover, there is 
    a non--zero
    efficiency for $\omega_1 >\frac{\Delta_1 \Gamma_1}{2 \varepsilon}$. At first glance, this is surprising because this is the non--pumping regime. However, there is a small $N_\text{pump}$ and also a small performed work which leads to $\eta_\text{pump}>0$.
    }
    \label{fig:2D_eff}
\end{figure}

Figure \ref{fig:2D_N} shows $N_{\text{pump}}$ as a function of the width $\Delta_1$ and the peak position $\omega_1$ of the spectral coupling--density $\Gamma_1(\omega,t)$ with $\Gamma_1=100\, V$, obtained from the exact solution (cf. Eq. \eqref{statnQD} and Appendix \ref{sec: exactAppendixstationaryQDn}). 
We find that $N_{\text{pump}}$ strongly depends on $\Delta_1$ and $\omega_1$. 
In particular, there is a line $\omega_1 = \frac{\Delta_1 \Gamma_1}{2 \varepsilon}$, above which pumping is strongly suppressed (region I). 
Below this line (regions II--IV), pumping occurs.
The dependence of $N_{\text{pump}}$ on $\Delta_1$ and $\omega_1$ can be qualitatively understood in terms of the energy $\omega_1$ of the first reaction--coordinate, the energy $\epsilon$ of the QD level and their coupling $\sqrt{\Gamma_1 \Delta_1/2}$.
For this, we diagonalize the subspace of reaction--coordinate 1 and QD level to get two eigenstates with energies
\begin{equation}\label{EpmQDRC}
    E^\text{QD-RC}_\pm = \frac{\varepsilon+\omega_1}{2}\pm \frac{1}{2}\sqrt{(\varepsilon-\omega_1)^2+ 2\Gamma_1 \Delta_1} \,. 
\end{equation}
For pumping, the state with the lower energy $E^\text{QD-RC}_-$ is relevant.
It must be occupied and must have a large overlap with the bare QD level in order to maximize $\left<n_{\text{stat},1}\right>$ and, therefore, also $N_\text{pump}$.

In the following, we distinguish four regimes, as depicted in Figure \ref{fig:RC_schematic_states}.
Region I is defined by $\omega_1>\frac{\Gamma_1 \Delta_1}{2 \varepsilon}$, for which the energy of the lower hybridized state lies above the chemical potential of bath 1, $E^\text{QD-RC}_- >0$.
In this case, the occupation of the hybridized level and, therefore, also $N_\text{pump}$ is suppressed%
\footnote{However, there will be a small but finite occupation because of off--resonant and higher--order tunneling processes.}.

A sizable occupation occurs only for $E^\text{QD-RC}_- < 0$, or, equivalently, $\frac{\Gamma_1 \Delta_1}{2 \varepsilon}> \omega_1$ (regions II--IV).
The largest overlap between the hybridized state and the bare QD level is achieved if the energy of reaction--coordinate 1 lies (in units of the coupling between the reaction coordinate and the QD level) well above the bare QD level, $\omega_1-\varepsilon \gg \sqrt{2 \Gamma_1 \Delta_1}$.
This defines region II, for which the pumped charge per cycle becomes maximal, $N_\text{pump}\approx 1$.
In regions III and IV, the energy $\omega_1$ of the reaction coordinate lies below the QD level $\varepsilon$.
For weak hybridization, $\sqrt{2 \Gamma_1 \Delta_1}\ll |\omega_1-\varepsilon|$ (region III), the overlap between the hybridized state and the QD is small, resulting in a small $N_\text{pump}$. 
Finally, for strong hybridization, $\sqrt{2 \Gamma_1 \Delta_1}\gg |\omega_1-\varepsilon|$ (region IV), the lower hybridized state is an antisymmetric superposition of the QD and the reaction coordinate, resulting in $N_\text{pump}\approx \frac{1}{2}$.

The pumped charge per cycle is only one number that characterizes the electron pump.
Another one is the energy efficiency $\eta_{\text{pump}}$.
It quantifies which percentage of the work performed during the coupling and decoupling procedure is converted into the chemical energy gain of the second bath.

The work performed on the device by changing the coupling between QD and baths at time $t$ is given by
the total energy change during the coupling or decoupling procedure. However, when as in our case only the coupling Hamiltonian changes abruptly, the state does not change and the work simplifies to
\begin{equation}\label{WC(t)}
    W_\mathrm{C}(t)=\left<H_\mathrm{C}(t+0^+) \right>-\left<H_\mathrm{C}(t-0^+) \right> \, ,
\end{equation} 
i.e., the change of the expectation value of the coupling Hamiltonian $H_\mathrm{C}$ at time $t$.
The total net performed work during the $\ell$-th cycle is, therefore, the sum $W_\mathrm{C}(\ell t_1+(\ell-1)t_2)+ W_\mathrm{C}(\ell t_1+\ell t_2)$ of the works performed during both switching processes. 
This yields, for the long--time limit, the energy efficiency
\begin{equation}
    \eta_\text{pump}=\lim_{\ell\rightarrow \infty}
    \frac{N_{\text{pump}}V}{ W_\mathrm{C}(\ell t_1+(\ell-1)t_2)+  W_\mathrm{C}(\ell t_1+\ell t_2)} \, ,
    \label{etapump}
\end{equation}
since $N_{\text{pump}}V$ is the chemical energy gain per pumping cycle (cf. Figure \ref{fig:2D_eff}).

While Eq.~\eqref{etapump} is valid for arbitrary $t_1$ and $t_2$, we show in Figure \ref{fig:2D_eff} the limit of slow pumping, $t_1,t_2 \to \infty$. 
In the wideband limit $\Delta_1 \rightarrow \infty$, excitations are generated all over the bath during the pumping process. This means that the performed work is mainly converted into excitations in the bath but not into the pumping of electrons. Therefore, $\eta_\text{pump}$ vanishes even though $N_\text{pump}$ can have a large value. 
In the opposite limit, $\Delta_1 \rightarrow 0$, effectively only one bath mode, the reaction coordinate, becomes excited during the pumping process. Here, the performed work splits half and half between excitations in the bath and pumping (cf. Appendix \ref{sec: exactsoMstep1_eta}), resulting in the maximally possible value $\eta_\text{pump}=\frac{1}{2}$ for slow pumping.

For finite values of $t_1$ and $t_2$, i.e., beyond slow pumping, pumping--efficiency values higher than $\frac{1}{2}$ are possible.
This we discuss in Section \ref{sec: improvefficency}.
The improved efficiency relies
on non--Markovian effects for finite $t_1$, $t_2$ and $\Gamma_2$. 
There, we focus on structured baths with small $\Delta_1$, because, as we have seen above, in that case the work performed by switching the couplings is converted more efficiently into the chemical energy gain of the second bath.
In addition, in this regime it is possible to study oscillations of $N_\text{pump}$ as an indicator for the impact of coherences between the QD and the baths.

%%%%%%%%%%%%%%%%%%%%%%%%%%%%%%%%%%%%%%%%%%%%%%%%%%
\subsection{Improved efficiency in generic pumping}
\label{sec: improvefficency}

\begin{figure} 
    \centering
    \includegraphics[width=0.9\linewidth]{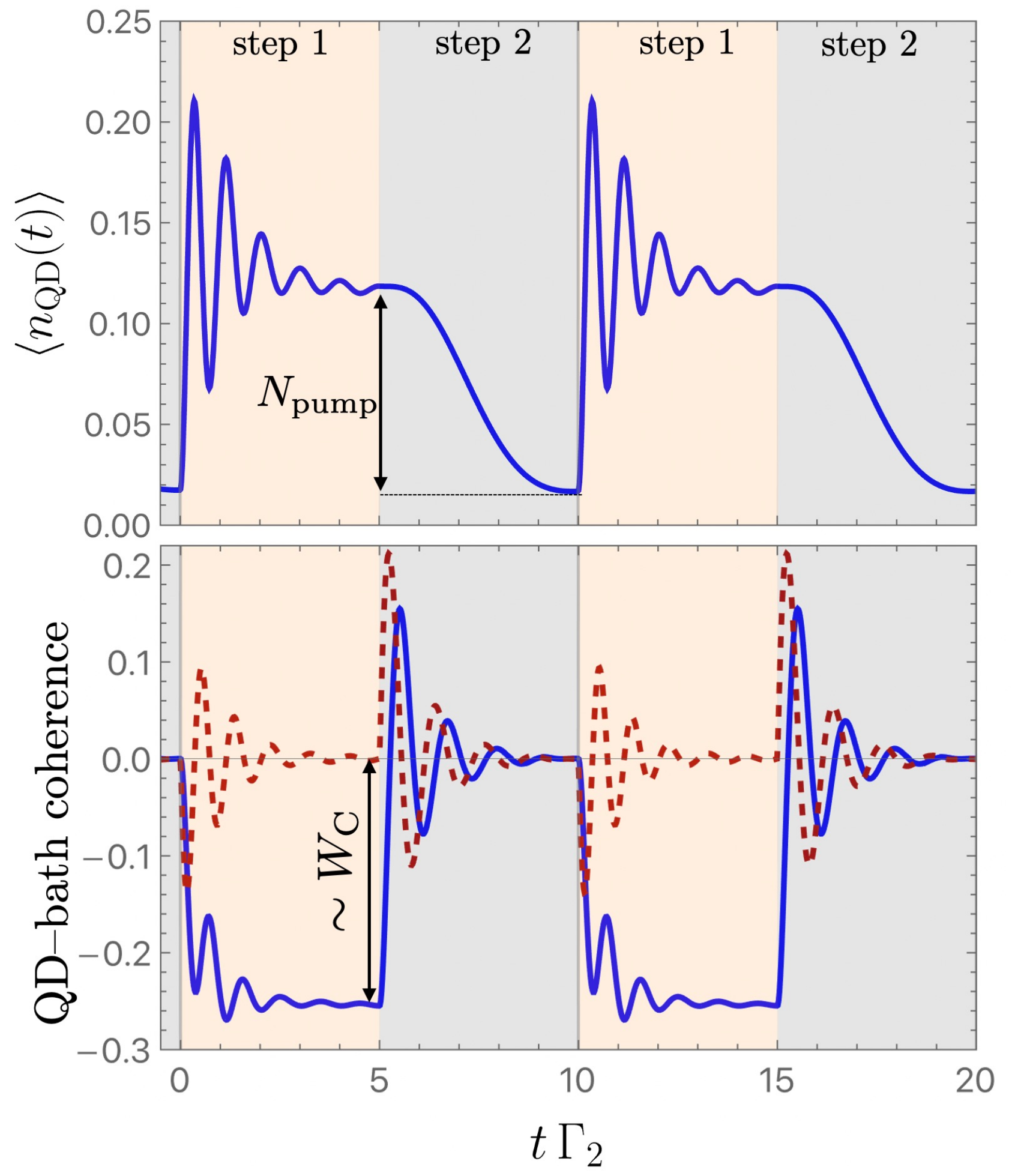}
    \caption{The Figures show the particle number of the QD (top) and the collective coherence $\left<r_1^\dagger c\right>$ between the QD and the first bath (cf. Eq. \eqref{rnud}) (bottom: blue--solid curve $\widehat{=}$ real part; red--dashed curve $\widehat{=}$ imaginary part) as functions of time $t$. Additionally, $N_\text{pump}$ and $W_\text{C}$ are shown in the plots. The durations of step 1 and step 2 are equal, $t_1=t_2=\frac{5}{\Gamma_2}=\frac{40}{V}$, and large compared to $\frac{1}{\Delta_1}$, $\frac{1}{\Delta_2}$, $\frac{1}{\Gamma_1}$ and $\frac{1}{\Gamma_2}$.
    Thus, the coherence $\left<r_1^\dagger c\right>$ vanishes almost completely at the end of step 2. However, this is different for shorter $t_2$ (cf. Figure \ref{fig:NBsp2}). 
    The other parameters are given by $\Gamma_1=\frac{5}{2}V$, $\Delta_1=\frac{V}{2}$, $\Gamma_2=\frac{1}{4}V$, $\Delta_2=\frac{1}{8}V$, $\varepsilon=\frac{5}{4}V$, $\omega_1=0$ and $\omega_2=\varepsilon$.}
    \label{fig:NBsp1}
\end{figure}
\begin{figure*}[!htb]
    \centering
    \includegraphics[width=\linewidth]{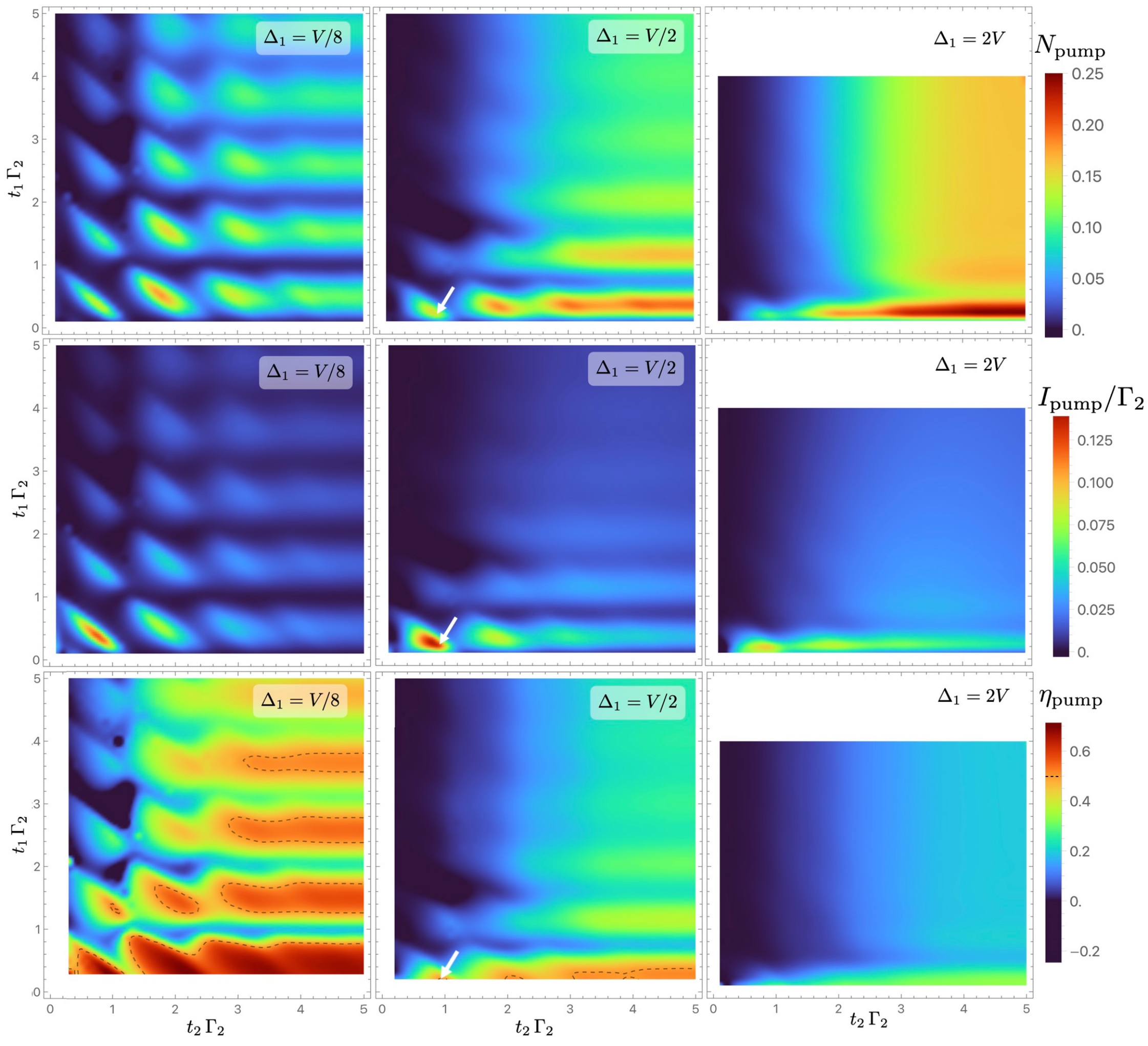}
    \caption{The colors show the averaged number of pumped electrons $N_\text{pump}$ (top--row), the current through the QD $I_\text{pump}$ (middle--row) and the energy efficiency $\eta_\text{pump}$ (bottom--row) as functions of the duration $t_2$ of step 2  and the duration $t_1$ of step 1. 
    All three quantities show non--Markovian oscillations. In particular, these effects lead to an increase of the energy efficiency, $\eta_\text{pump}$, above the upper bound, $\eta_\text{pump}=\frac{1}{2}$, of the slow pumping regime, indicated by dashed--black contour lines (two left figures of the bottom--row).
    The coupling with the first bath is given by $\Gamma_1=\frac{5}{2}V$ and the coupling with the second bath is given by $\Gamma_2= \frac{V}{8}$. The peak position of the {spectral coupling--density} of the first bath is given by $\omega_1=0$ and the width varies with each column ($\Delta_1=\frac{V}{8}$, $\Delta_1=\frac{V}{2}$, $\Delta_1=2V$). In contrast to this, the peak position of the {spectral coupling--density} of the second bath is located at the QD--orbital energy $\omega_2= \varepsilon= \frac{5}{4}V$ and the width is given by $\Delta_2 = \frac{V}{4}$. The white arrow corresponds to the situation shown in Figure \ref{fig:NBsp2}.
    For numerical reasons, we compute $N_\text{pump}\approx N_{\text{pump},10}$, $I_\text{pump}\approx I_{\text{pump},10}$ and $\eta_\text{pump}\approx\eta_{\text{pump},10}$ for the tenth cycle and assume that this is identical to the limit cycle.  
    However, this assumption is not valid for $\eta_\text{pump}$, small $t_1+t_2$ and small $\Delta_1$ in particular. Therefore, the plots for $\eta_\text{pump}$ have an adjusted and restricted plot range. For $\Delta_1=2 V$, there is a restricted plot range for all quantities because of computational reasons.  }
    \label{fig:Npump}
\end{figure*} 
\begin{figure} 
    \centering
    \includegraphics[width=0.9\linewidth]{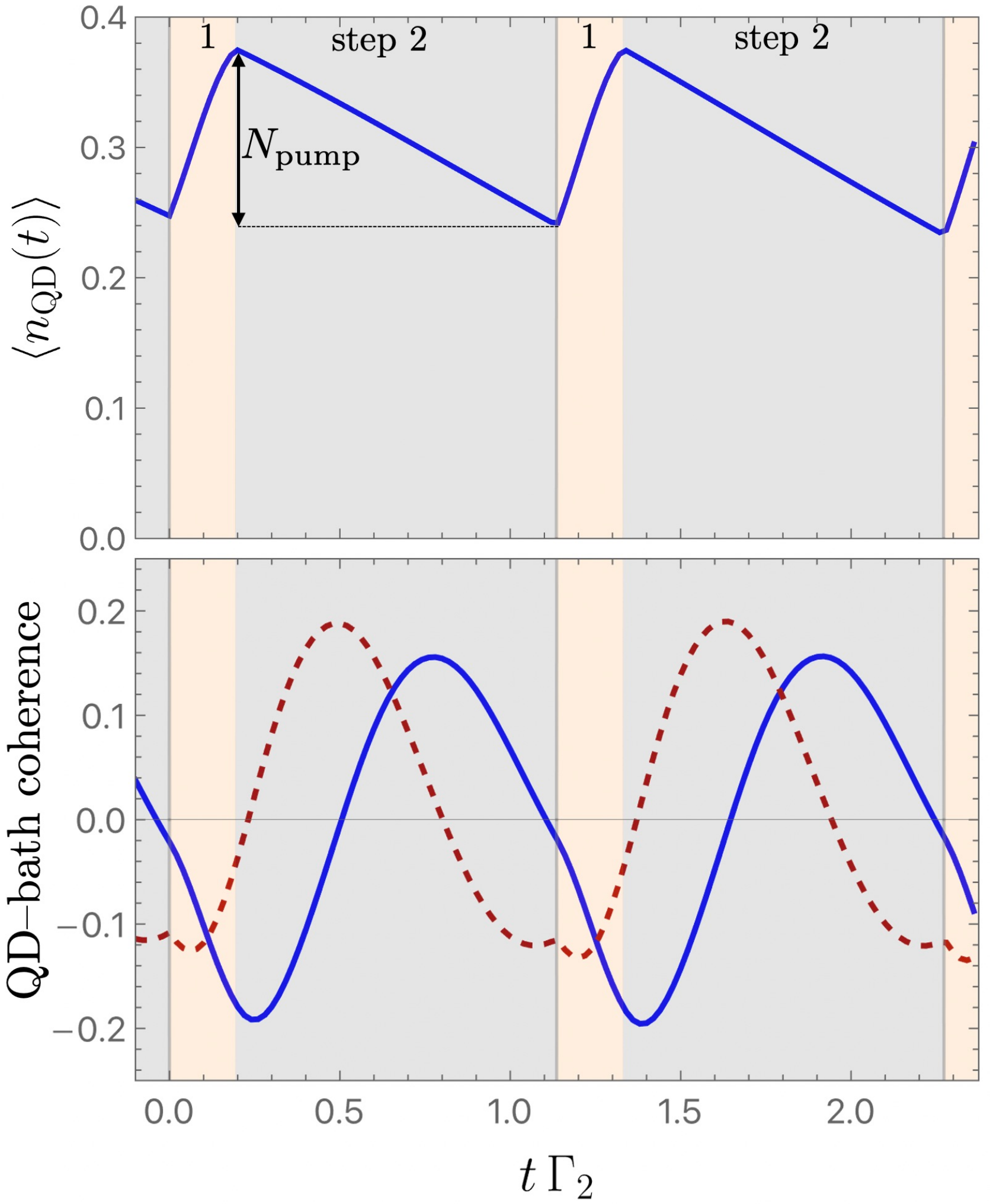}
    \caption{The Figures show the particle number of the QD (top) and the coherence $\left<r_1^\dagger c\right>$ (cf. Eq. \eqref{rnud}) (bottom: blue--solid curve $\widehat{=}$ real part; red--dashed curve $\widehat{=}$ imaginary part) as functions of time $t$. The durations of step 1 and step 2 are $t_1=\frac{\pi}{2 \sqrt{\varepsilon^2+2 \Gamma_1 \Delta_1 }}\neq t_2= \frac{3 \pi}{2 \varepsilon}$ and correspond to the white arrow in Figure \ref{fig:Npump}. In contrast to Figure \ref{fig:NBsp1}, $t_2$ is so short that $\left<r_1^\dagger c\right>$ does not vanish during step 2. This leads to an increase of the current during step 1. In addition, $t_1$ is chosen such that the non--Markovian oscillations of the particle number lead to an improved value of $N_\text{pump}$. 
    The other parameters are given by $\Gamma_1=\frac{5}{2}V$, $\Delta_1=\frac{V}{2}$, $\Gamma_2=\frac{V}{4}$, $\Delta_2=\frac{V}{8}$, $\varepsilon=\frac{5}{4}V$, $\omega_1=0$ and $\omega_2=\varepsilon$.}
    \label{fig:NBsp2}
\end{figure}

In the slow--pumping regime ($t_1,\,t_2 \rightarrow \infty$), described in Section \ref{sec:slowpumping}, the coherence between the QD and the bath to which it is momentarily coupled has reached its stationary limit, and the coherence with the other bath has fully decayed.
In the present section, we study the more complicated situation that the step durations $t_1$ and $t_2$ are kept finite, for which the time dependence of the coherence becomes important.
For the calculations we use the exact solution introduced in Appendix \ref{sec: nonhermitianOperatorAppendix}.

Strong  
non--Markovian effects such as oscillations of the QD occupation may appear%
\footnote{In general, oscillations can occur in Markovian systems, but for a single QD, oscillations in the occupation can be directly linked to information back--flow and thus to non-Markovian effects \cite{breuerFoundationsMeasuresQuantum2012}.}.
As one result, one may achieve an improved energy efficiency $\eta_\text{pump}$ for pumping. 

The coherence between the QD and, say, bath 1 is quantitatively characterized by the expectation value $\left<r_1^\dagger c\right>$ involving one operator for the QD and one for reaction--coordinate 1.
In the presence of a tunneling coupling (i.e., during step 1), the real part of the coherence is related to the coupling energy
\begin{equation}
    \left<H_\mathrm{C}(t)\right>=2 \sqrt{\frac{\Gamma_1 \Delta_1}{2}} \, \mathrm{Re}\left<r_1^\dagger c\right> \, ,
\end{equation}
while the imaginary part describes the current from bath 1 to the QD via
\begin{equation}\label{I1QD}
    I_{1\rightarrow \text{QD}}= - 2 \sqrt{\frac{\Gamma_1 \Delta_1}{2}} \mathrm{Im}\left<r_1^\dagger c\right> \, .
\end{equation}

In order to reach high energy efficiency, we use parameters $\omega_1$ and $\Delta_1$ in the lower left region of Figure \ref{fig:2D_eff}.
Moreover, we choose $\Gamma_2 \ll \Gamma_1$, $\omega_2=\varepsilon$  and $\Delta_2 < \varepsilon- V$ to suppress unwanted tunneling from the second bath to the QD. 

The top panel of Figure \ref{fig:NBsp1} displays
the QD occupation number as a function of time. 
The orange and gray areas indicate step 1 and step 2, respectively. 
During step 1, the electrons coherently oscillate with frequency $\sqrt{\varepsilon^2+2 \Gamma_1 \Delta_1 }$ between bath 1 and the QD for a short time.
This is a consequence of the peaked bath structure. 
The amplitude of the oscillation decays with time, such that the QD occupation approaches the stationary QD occupation $\left<n_\text{stat,1}\right>$ (given by Eq. \eqref{statnQD} in Section \ref{sec:slowpumping}).
For finite $t_1$, the values of $N_\text{pump}$, $I_\text{pump}$, and also $\eta_\text{pump}$ depend on whether 
the end of step 1 corresponds to a high or a low point of this oscillation.

The bottom panel of Figure \ref{fig:NBsp1} displays both the real and the imaginary part of the coherence between QD and bath 1 as a function of time.
It is interesting to note that the coherence is not only present during step 1, in which the QD is coupled to bath 1.
There is finite (but decaying) coherence also during step 2.
Since there is no tunneling coupling to bath 1 during step 2, the real and imaginary part of the QD--bath 1 coherence are not associated with a coupling energy or a charge transfer.
Nevertheless, the decaying oscillations during step 2 do have an impact on pumping if $t_2$ is sufficiently small such that at the end of step 2 the coherence between QD and bath 1 has not yet fully vanished.
Depending on whether the imaginary part has a high or low point at the end of step 2, a current flows directly from or to the QD at the beginning of the following step 1. 

The influence of finite values of $t_1$ and $t_2$ is illustrated in Figure \ref{fig:Npump}, which shows, at small times and small enough $\Delta_1$, pronounced oscillations of $N_\text{pump}$, $I_\text{pump}$, and $\eta_\text{pump}$ as a function of $t_1$ and $t_2$.
Figure \ref{fig:NBsp2} shows an optimal situation where $t_1$ is chosen such that the first oscillation peak of the particle number corresponds to the end of step 1 and $t_2$ is chosen so that the imaginary part of the coherence between the first bath and the QD has a low point at the end of step 2. These parameters correspond to an enhanced $I_\text{pump}$, as indicated with a white arrow in Figure \ref{fig:Npump}. 

With increasing $\Delta_1$, the oscillations of the coherences between the baths and the QD decay faster and, finally, become invisible in $N_\text{pump}$, $I_\text{pump}$, and $\eta_\text{pump}$.
A large energy efficiency $\eta_\text{pump}$ is achieved for small $\Delta_1$, for which oscillations are pronounced, by properly choosing $t_2$.
On the one hand, the efficiency also increases as $N_\text{pump}$ increases due to a large negative value of the imaginary part of the coherence at the end of step 2. 
On the other hand, efficiency increases as the work performed decreases due to a large positive value of the real part of the coherence at the end of step 2.
This interplay leads to the relative shift between the maxima of $N_\text{pump}$ and $\eta_\text{pump}$, as shown in Figure \ref{fig:Npump}, the latter partially exceeding $\frac{1}{2}$. For further discussion, in particular on how the energy efficiency is limited by bath excitations, see Appendix \ref{sec:appExcitaionsBath}. 

As a result, non--Markovian effects can increase the efficiency with which the work performed by coupling and decoupling is converted into the chemical energy gain of the second bath.
Moreover, the presence of oscillations of $N_\text{pump}$, $I_\text{pump}$ and $\eta_\text{pump}$ as functions of $t_1$ and $t_2$ is an indicator that coherences and the coupling energy play an important role.
{For this, it is essential that $t_1$ and $t_2$ are chosen accurately, where the tolerance range is smaller than $1/\sqrt{\varepsilon^2+2 \Gamma_1 \Delta_1 }$ for $t_1$ and smaller than $1/\varepsilon$ for $t_2$. }

\newpage
\section{scheme 2: measurement induced pumping}\label{sec:scheme2}
Similar to the first pumping scheme, also the second one relies on using the coupling energy between quantum dot (QD) and baths through decoherence processes.
In pumping scheme 1, the decoherence is triggered by decoupling the QD from the bath.
This requires a time--dependent variation of the tunnel couplings.
In pumping scheme 2, decoherence is invoked by measuring the QD occupation.

\begin{figure}
    \centering
    \includegraphics[width=\linewidth]{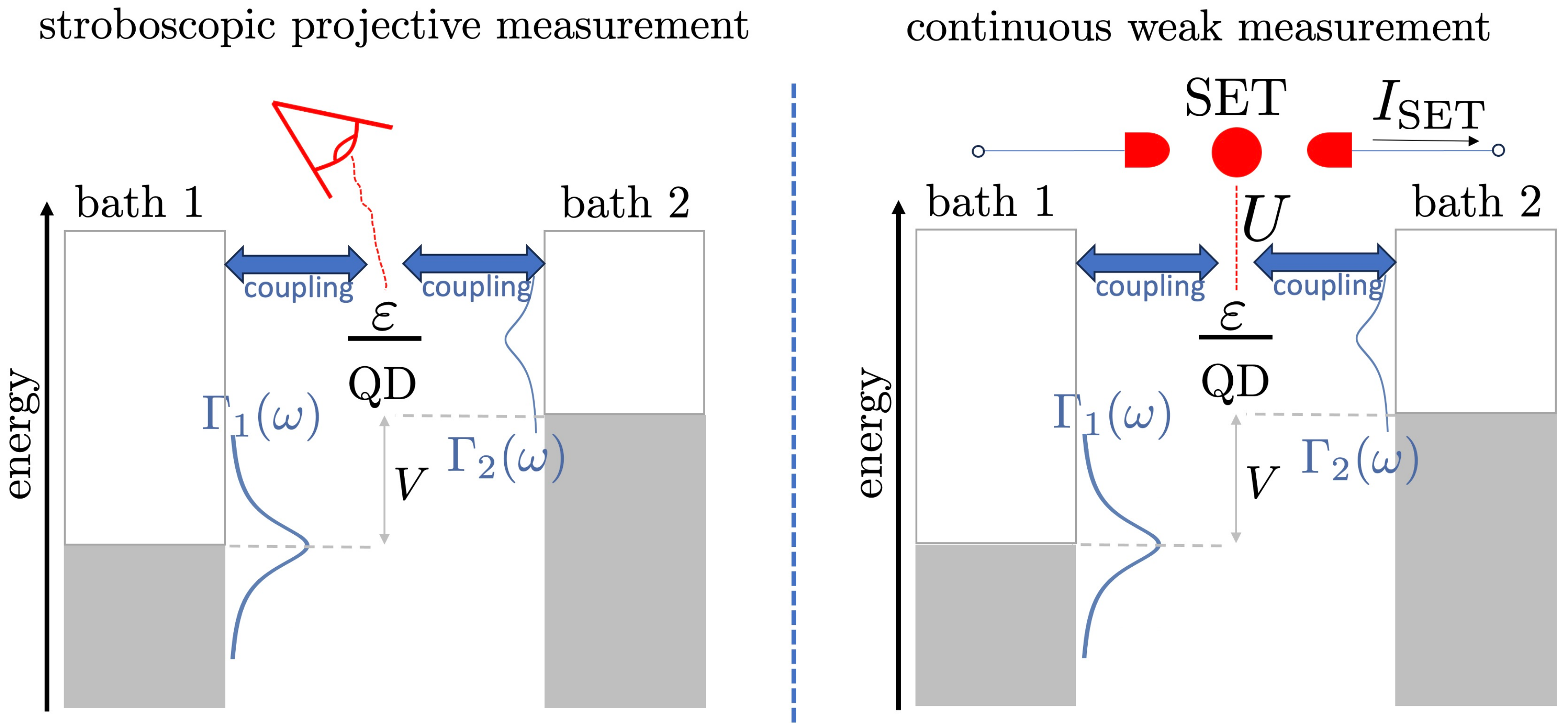}
    \caption{Schematic visualization of the functionality of the second pumping scheme discussed in Section \ref{sec:scheme2}: In contrast to the electron pump shown in Figure \ref{fig:Schaubild2}, here both baths are always coupled to the QD. Stroboscopic projective measurements (left) or a continuous and weak measurement via a single--electron transistor (SET) with interaction $U$ and current $I_\SET$ (right) are applied to the QD. For appropriately chosen spectral densities $\Gamma_1(\omega)$, $\Gamma_2(\omega)$ the measurements induce electron transport from bath 1 to bath 2.}
    \label{fig:AZ_Schaubild}
\end{figure}
This measurement--induced pumping mechanism bears similarities to the anti--Zeno effect \cite{AZ1,AZ2,AZ_fujii2010anti,QZuAZenoQD} and to a contemporaneous study for a similar device in \cite{sanchez_Making_2026}. 

In contrast to the first pumping scheme, we will now consider a setup in which both baths are always coupled to the QD ($g_\nu(t)\equiv 1, \nu=1, 2$ in Eq. \eqref{Gamma_nu(w,t)}), where the structure of the baths determines a preferred direction of the tunneling. 
Decoherence occurs by coupling a charge detector to the QD.
This could be done in various ways.
Here, we discuss two limiting cases, see Figure \ref{fig:AZ_Schaubild}.
In the first scenario, a strong, i.e., projective measurement is performed in a stroboscopic manner --- see left panel of Figure \ref{fig:AZ_Schaubild}.
This scheme, discussed in Section \ref{sec:projecMea}, has the conceptual advantage that the decoherence occurs at well--defined, externally controlled instants in time.
The second scenario, described in Section \ref{sec:SET}, relies on a continuous and weak measurement ---  see right panel of Figure \ref{fig:AZ_Schaubild}.
Such steady--state situation may be easier to implement experimentally.

\FloatBarrier

\subsection{Stroboscopic projective measurement} \label{sec:projecMea}

\begin{figure}
    \centering
    \includegraphics[width=\linewidth]{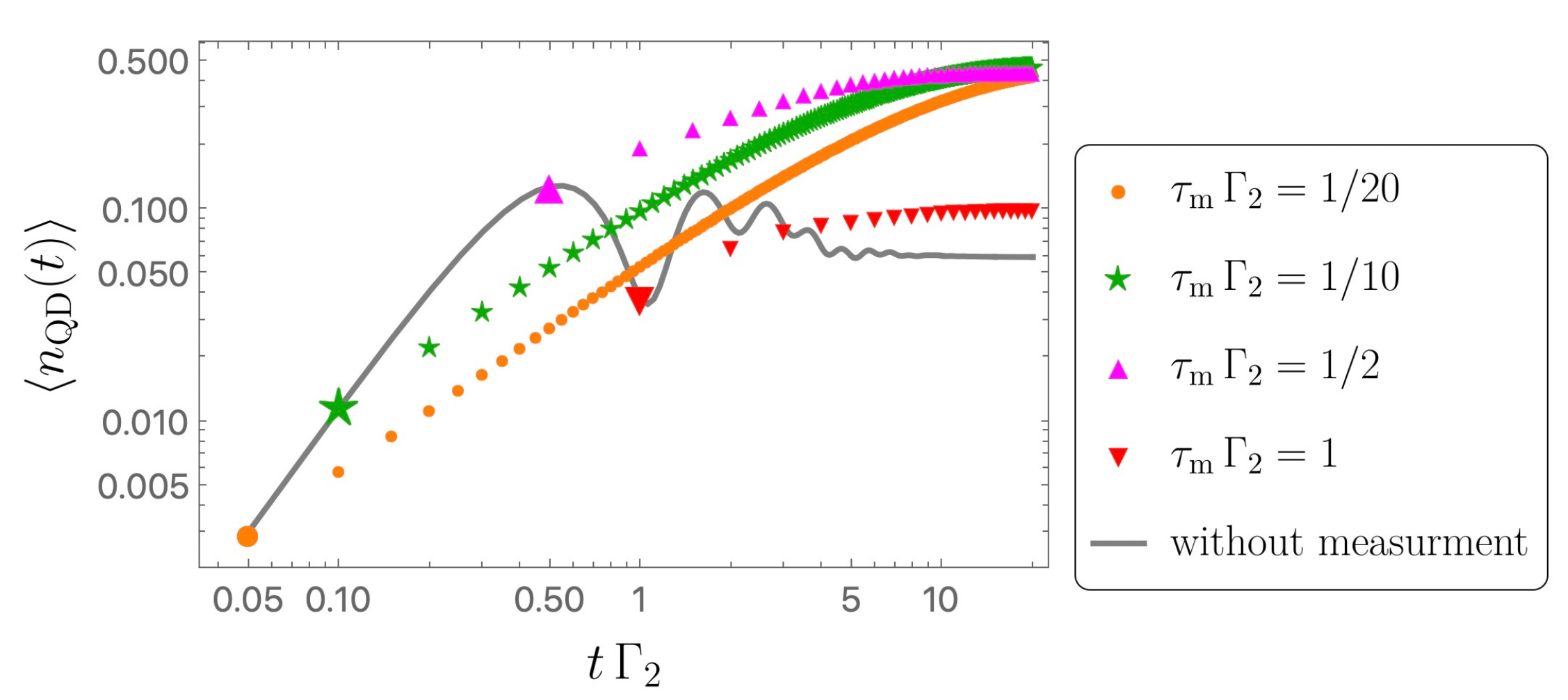}
    \caption{The expectation value of the QD occupation $\left<n_\text{QD}\right>$ as a function of the time $t$ that has elapsed since the QD was in the empty initial state (double logarithmic scale): The solid curve shows $\left<n_\text{QD}\right>$ without repeated measurements, whereas the symbols show $\left<n_\text{QD}\right>$ for periodically repeated measurements  with $\tau_\text{m}$ and averaged over all results of the measurements. For $t\rightarrow \infty$, $\left<n_\text{QD}\right>$ becomes stationary with as well as without measurements. The stationary $\left<n_\text{QD}\right>$ with measurements are larger than without measurements.
    In contrast, the rapidity with which $\left<n_\text{QD}\right>$ converges toward its stationary value is slowed down for short measurement intervals $\tau_\text{m}$ (Zeno--effect \cite{misra1977zeno}). For the calculations, we use the exact solution discussed in Appendix \ref{sec:exactwithmeasurement}. The parameters are given by $\omega_1=0$, $\omega_2=\varepsilon=\frac{5}{4}V$, $\Gamma_1=\frac{5}{2}V$, $\Gamma_2=\frac{V}{4}$ and $\Delta_1=\Delta_2=\frac{V}{8}$. }
    \label{fig:AZ_n_QD(t)}
\end{figure}

If both baths stay coupled to the QD and no measurements are performed, the applied bias voltage $V$ will drive, via higher--order tunneling processes, a constant current% 
\footnote{This is related to the strict positivity of the entropy production rate, $ \text{d}S/\text{d}t = \beta (\mu_1 -\mu_2) I_\text{stat} > 0$, \cite{nenciu_independent_2007}.}
$I_\text{stat}$ from bath 2 to bath 1.
This can, however, change in presence of measurements.
A projective measurement of the QD particle number
and averaging over all measurement outcomes 
modifies the density matrix by only deleting the coherences between the QD and the baths%
\footnote{However, the coherences between the baths and the coherences inside the baths remain unaffected.}
(cf. Appendix \ref{sec:exactwithmeasurement} for the exact description).
At early times after the measurement, the electrons may enter the QD even if $\varepsilon>V$, involving the rebuild of coherences between the baths and the QD. However, these processes are suppressed for later times when the coherences become stationary. 
\begin{figure}
    \centering
    \includegraphics[width=0.8\linewidth]{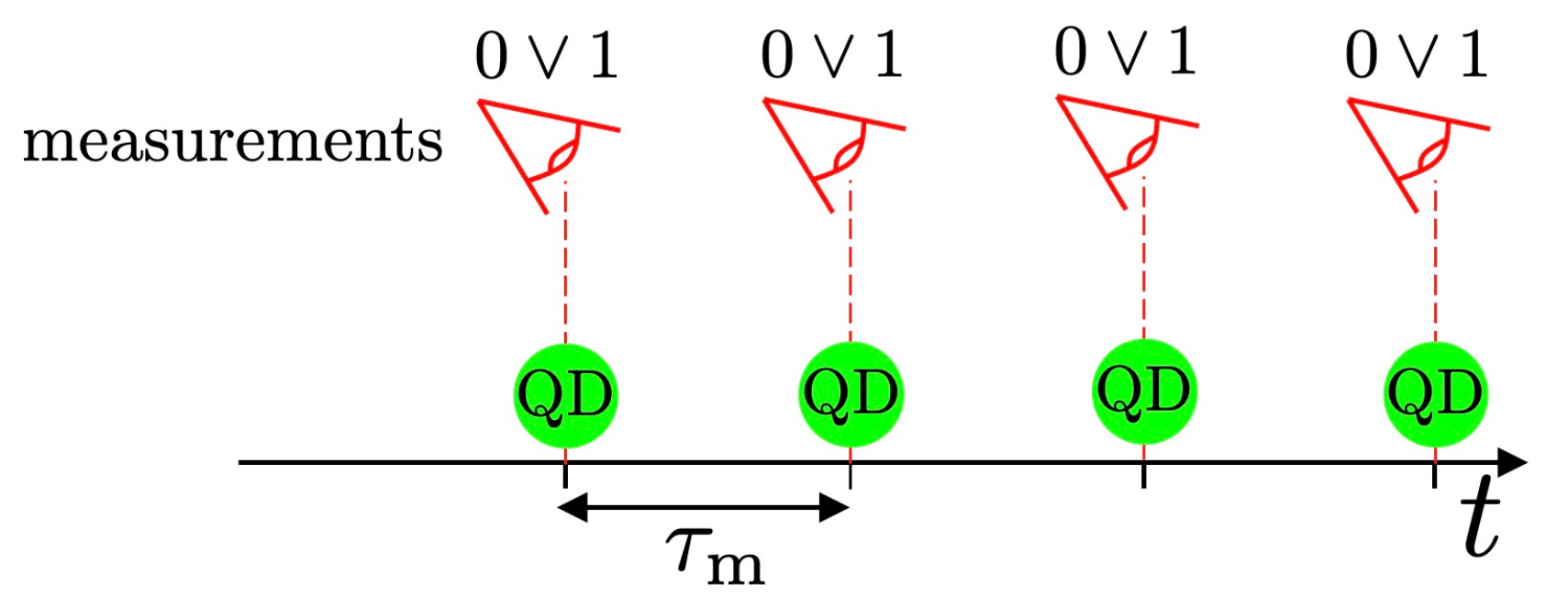}
    \caption{Schematic visualization of the measurement scheme discussed in Section \ref{sec:projecMea} as a function of time $t$. The measurements of the particle number of the QD ($0$ or $1$) are repeated stroboscopically with the time interval $\tau_\text{m}$.}
    \label{fig:AZ_tau}
\end{figure}
For measurements that are repeated stroboscopically with time interval $\tau_\text{m}$ (cf. Figure \ref{fig:AZ_tau}), the QD can gain energy from the system--bath coupling for short times after each measurement \cite{Teff} (deletion of the  coherences) which results in a higher final occupation number of the QD than without the repeated measurements (cf. Figure \ref{fig:AZ_n_QD(t)}). This effect is similar to the anti--Zeno effect \cite{AZ1,AZ2,AZ_fujii2010anti,QZuAZenoQD}%
\footnote{But also the Zeno effect \cite{misra1977zeno} is present for short $\tau_\text{m}$ and $t$ (cf. Figure \ref{fig:AZ_n_QD(t)}).}.
%%%%%%%%%%%%%%%%%%

%%%%%%%%%%%%
A proper choice of the spectral densities%
\footnote{
One possibility of generating structured baths is to add additional QDs between the original QD and the baths, as discussed in Section \ref{sec:RC} for the reaction--coordinate mapping.}
$\Gamma_1(\omega)$ and $\Gamma_2(\omega)$
(with strong coupling to occupied states of bath 1 and coupling to mainly unoccupied states of bath 2, respectively) may lead to an electric current flowing against $V$: from bath 1 to bath 2.
\begin{figure}
    \centering
    \includegraphics[width=\linewidth]{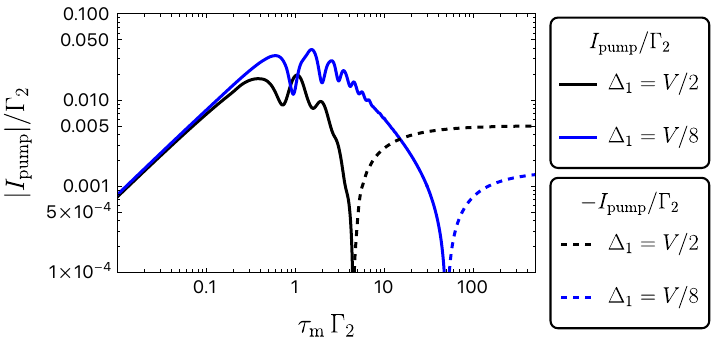}
    \caption{The averaged current $I_\text{pump}$, induced by periodically repeated measurements, is shown as a function of the measurement interval $\tau_\text{m}$ (in units of the coupling strength $\Gamma_2$ between the QD and bath 2)
    for two different values $\Delta_1=\frac{V}{8}$ (blue) and $\Delta_1=\frac{V}{2}$ (black)
    in a double logarithmic scale. It increases linearly for short $\tau_\text{m}$ (Zeno regime), 
    equally for both values of $\Delta_1$, since the weak coupling between the QD and the second bath becomes the common limiting factor.
    Later, there appear oscillations corresponding to electrons oscillating between the first bath and the QD. For large $\tau_\text{m}$, the averaged current decays as $1/\tau_\text{m}$ against the stationary current $I_\text{stat}<0$ without measurements, where a negative $I_\text{pump}$ is indicated by a dashed curve and its absolute value increases with increasing $\Delta_\nu$. 
    The other parameters are $\omega_1=0$, $\omega_2=\varepsilon=\frac{5}{4}V$, $\Gamma_1=\frac{5}{2}V$, $\Gamma_2=\frac{V}{4}$, $\Delta_2=\frac{V}{8}$. 
    The exact solution discussed in Appendix \ref{sec:exactwithmeasurement} is used for the calculations. 
    }
    \label{fig:AZ_I_pump}
\end{figure}
The required energy is obtained from the measurement process%
\footnote{The energy efficiency $\eta_\text{pump}$ can be calculated similarly to Eq.~\eqref{etapump}. However, in reality, infinite work ($\eta_\text{pump}\rightarrow 0$) needs to be performed to realize an ideal projective measurement \cite{guryanovaIdealProjectiveMeasurements2020}.}.
Figure \ref{fig:AZ_I_pump} shows the current averaged over the pumping cycle
\begin{equation}
\begin{split}
    I_\text{pump}&=\frac{1}{\tau_\text{m}}\lim_{\ell\rightarrow \infty} \int_{(\ell-1)\tau_\text{m}}^{\ell\,\tau_\text{m}} \text{d}t\,I_{1\rightarrow \text{QD}}(t)\\
    &=-\frac{1}{\tau_\text{m}}\lim_{\ell\rightarrow \infty} \int_{(\ell-1)\tau_\text{m}}^{\ell\,\tau_\text{m}}\text{d}t\,I_{2\rightarrow \text{QD}}(t)
\end{split}
\end{equation}
(cf. Eq. \eqref{I1QD}) as a function of the measurement interval $\tau_\text{m}$.
In the limit of infinitely frequent measurements ($\tau_\text{m}\rightarrow 0$), $I_\text{pump}$ vanishes because the measurements always delete the coherences between the QD and the baths, thus stopping any dynamics, which is known as the Zeno effect \cite{misra1977zeno}. 
Specifically, the current between the baths and the QD increases linearly in time $t$ starting from zero directly after each measurement. 
As a consequence, also the averaged current $I_\text{pump}$ increases linearly with the interval $\tau_\text{m}$ for short $\tau_\text{m}$. 
However, the current between the baths and the QD deviates from the linear behavior for later times, when oscillations between the baths and the QD become relevant. This can be observed as oscillations in the averaged current (cf. Figure \ref{fig:AZ_I_pump})%
\footnote{These oscillations are similar to the oscillations (as a function of $t_1$) for the coupling/decoupling procedure (cf. Section \ref{sec: improvefficency}).}.
After the energy injected by the measurement process is used up, there is no pumping against the bias anymore in that cycle. The averaged current then decays as $1/\tau_\text{m}$ against the negative stationary current $I_\text{stat}<0$ from bath 2 to bath 1. 
The impact of the higher--order tunneling processes enabling $I_\text{stat}$  increases as $\Delta_\nu$ increases (cf. black and blue curves in Figure \ref{fig:AZ_I_pump}).

\subsection{Continuous weak measurement} 
\label{sec:SET}

The realization of a stroboscopic strong measurement scheme requires a higher degree of control than just permanently coupling a charge detector to the QD, which continuously performs a weak measurement.
As we show in this section, a continuous weak measurement can be sufficient to establish measurement--induced pumping.
For this, we consider 
a single--electron transistor (SET) that is capacitively coupled to the QD. 
Such setups are sometimes \cite{schoelkopfRadioFrequencySingleElectronTransistor1998,barthelFastSensingDoubledot2010} used to measure the occupation of the QD and the shot noise of the additional channel can induce drag or ratchet currents \cite{Ratchet2015,bischoff_Measurement_2015,drag_lim2018engineering} through the QD.
This is modeled by the total Hamiltonian
\begin{multline}
    \label{tildeHtot}
     H'=H+U n_\SET\,c^\dagger c +\sum_{\nu,k}\left(\gamma_\SET \,b_{\nu,k}^\dagger \,c_\SET + \text{h.c.}\right) \\+\sum_{\nu,k} \varepsilon_{\nu,k} b^\dagger_{\nu,k} b_{\nu,k}\, ,
\end{multline}
where we add an additional QD with fermionic annihilation and creation operators $c_\SET/c^\dagger_\SET$ (and $n_\SET=c^\dagger_\SET c_\SET$) to the Hamiltonian $H$ (cf. Eq. \eqref{Htot} and Eq. \eqref{Gamma_nu(w,t)} with $g_\nu(t)=1$). This QD interacts with the original QD via Coulomb interaction $U$ and is connected to two baths $\nu=\text{source}, \text{drain}$ {(of the SET)} with fermionic annihilation/creation operators $b_{\nu,k}/b_{\nu,k}^\dagger$. 
For simplicity, we have set the energy level of the SET to zero, since only the relative differences to the chemical potentials of the source and the drain are relevant.
The chemical potentials of the baths are chosen as $\mu_\text{source}=U/2$, $\mu_\text{drain}<0$, so that the lowest--order tunneling processes only permit a current between the source and the drain if the original QD is unoccupied. 
Therefore, the current flowing through the SET can be interpreted as a continuous measurement of the particle number of the original QD. Compared to $\tau_\text{m}$ for projective measurements, here, the average time $\tau_{e}$ between two consecutive electrons flowing through the SET indicates the strength of the measurement. That time is given by
the inverse of the average current%
\footnote{
The average time between two consecutive electrons flowing through the SET can be defined by $\displaystyle\tau_{e}\equiv\lim_{t\rightarrow \infty } t/N_\SET(t)=1/I_\SET$,
where $N_\SET(t)$ is the number of electrons flowing through the SET during the time interval $t$.}
flowing through the SET, $\tau_{e}=1/I_\SET$.

\begin{figure}
    \centering
    \includegraphics[width=\linewidth]{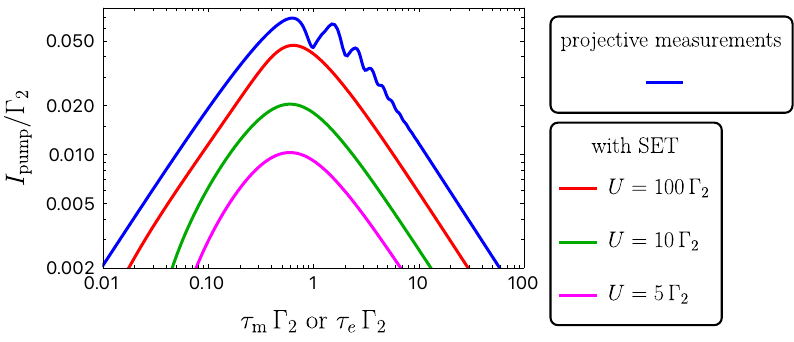}
    
    \caption{The averaged current $I_\text{pump}$ induced by periodically repeated measurements (blue) or the SET (other colors) is shown as a function of the measurement interval $\tau_\text{m}$ (blue) or the average time between two consecutive electrons flowing through the SET $\tau_{e}=1/I_\SET$ (other colors) in units of the coupling strength $\Gamma_2$ between the QD and bath 2 (double logarithmic scale). Here, the results are calculated with the Redfield equation in combination with the reaction--coordinate mapping (cf. Appendix \ref{sec:AppRedfield}). The Redfield solution for the periodically repeated measurements differs from the exact solution discussed in Figure \ref{fig:AZ_I_pump}. In particular, the Redfield equation prevents off--resonant tunneling processes from bath 2 to bath 1 that are present for the exact solution and cause negative currents in Figure \ref{fig:AZ_I_pump}. Thus, we are in a regime where the Redfield solution cannot be applied quantitatively, but still shows qualitative agreement with the exact solution. The currents for the SET show a similar behavior as for the projective measurements but show no oscillations and vanish for $U\rightarrow 0$.
    The parameters are given by $\omega_1=0$, $\omega_2=\varepsilon=\frac{5}{4}V$, $\Gamma_1=\frac{5}{2}V$, $\Gamma_2=\frac{V}{4}$, $\Delta_2=\frac{V}{8}$ and $\Delta_1=\frac{V}{8}$.}
    \label{fig:SET_Ipump}
\end{figure}
Figure \ref{fig:SET_Ipump} shows the averaged current $I_\text{pump}$ induced by projective measurements (blue {curve}) and the SET (other colors) as a function of $\tau_\text{m}$ or $\tau_{e}=1/I_\SET$. For calculations we use the Redfield equation in combination with fermionic reaction--coordinate mapping (cf. Appendix \ref{sec:AppRedfield}). The curves corresponding to the situation with SET are similar to those of the projective measurements. 
The reason is the noise from the electron tunneling through the SET which causes the QD energy level to fluctuate between $\varepsilon$ and $\varepsilon+U$. It is known \cite{Gurvitz_2019} that random fluctuations of the energy level of the QD lead to decoherence between the QD and the baths, which has a similar effect as the deletion of the coherences after the projective measurement%
\footnote{In contrast to \cite{Gurvitz_2019}, we consider a situation where fluctuations in the energy level of the QD are not completely random because, in our case, the occupations of the QD and the SET influence each other.

Our situation is similar to  \cite{sanchez_Making_2026} or to the drag current discussed in \cite{drag_lim2018engineering}, where the main effect results from shot noise--induced decoherence.
}.
However, the decay of the coherence, which corresponds to a fluctuation, is not instantaneous for finite $U$, so the measurement induced $I_\text{pump}$ is suppressed compared to the projective measurements and vanishes as $\sim U^2$ for small $U$. 
In addition, the capacitive coupling $U$ 
blocks the current flow in the case when the SET is occupied
and therefore 
$I_\text{pump}$ stays smaller than for the projective measurements,
even in the limit $U\rightarrow \infty$.
Furthermore, the time interval between the entry of two electrons into the SET is not constant, so the oscillations of $I_\text{pump}$ average out,
contrary to the periodically repeated projective measurements. 

\section{Comparison of the two pumping schemes}

\label{sec:scheme2vsCoUDeCou}
In the above sections, we have seen that it is possible to pump electrons by using a coupling/decoupling procedure (cf. Section \ref{sec:scheme1}) as well as by performing measurements of the QD occupation (cf. Section \ref{sec:scheme2}).  
In both cases, the pumping processes are related to externally induced decoherence operations (cf. Figure \ref{fig:decoherenceoperation}). Although the exact realization of these operations differ from each other, it is possible to compare the results of the two pumping schemes not only qualitatively but also quantitatively.

\begin{figure}
    \centering
    \includegraphics[width=\linewidth]{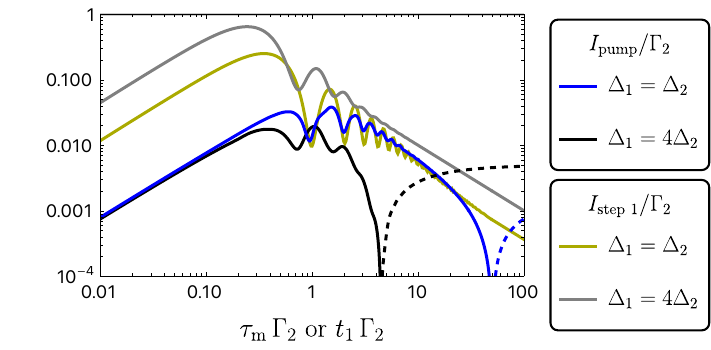}    
    \caption{The averaged current $I_\text{pump}$ induced by projective measurements (cf. Figure \ref{fig:AZ_I_pump}) and the current $I_\text{step 1}$ (cf. Eq. \eqref{Istep1} and Section \ref{sec: improvefficency}) as functions of the measuring interval $\tau_\text{m}$ or the duration $t_1$ of step 1 (in units of the coupling strength $\Gamma_2$ between the QD and bath 2). Negative values are indicated by dashed {curves}. The parameters are given by $t_2\rightarrow \infty$, $\omega_1=0$, $\omega_2=\varepsilon=\frac{5}{4}V$, $\Gamma_1=\frac{5}{2}V$, $\Gamma_2=\frac{V}{4}$, $\Delta_2=\frac{V}{8}$ and $\Delta_1=\frac{V}{8}$ or $\Delta_1=\frac{V}{2}$.}
    \label{fig:AZvsCou}
\end{figure}

For this, we consider the coupling/decoupling scheme (scheme 1) in the limit $t_2 \rightarrow \infty$. Then, there will be a total decay of the  coherences between the QD and bath 1 during step 2. Consequently, 
at the beginning of step 1 there will be no coherences between the QD and bath 1, similar to the situation after a projective measurement.
In contrast, after the measurement (scheme 2), there always remain some coherences between different bath modes which influence the dynamics (cf. Appendix \ref{sec: appendixmeasurementwideband}).
Figure \ref{fig:AZvsCou} compares the averaged current $I_\text{pump}$ for the projective measurements with the averaged current
\begin{equation}\label{Istep1}
    I_\text{step 1}=\frac{N_\text{pump}}{t_1}
\end{equation}
during step 1 for the coupling/decoupling procedure. The two currents show qualitatively similar behavior for both pumping schemes. Moreover, for $\Delta_1=\Delta_2$ (blue and dark yellow), the frequency and phase of the oscillations coincide quantitatively for moderate durations ($\tau_\text{m},\, t_1 \in [1/\Gamma_2,10/\Gamma_2]$)%
\footnote{For $\Delta_1\neq \Delta_2$, $ I_\text{step 1}$ scales with $\Delta_1/\Delta_2$.}. 
For longer durations (up to $t_1,\, \tau_\text{m}=30/\Gamma_2$), 
the corresponding currents are also quantitatively the same.
However, for the pumping scheme with measurements, there is always a coupling between both baths and the QD, 
by which $I_\text{pump}$ becomes negative for long durations,
while in the coupling/decoupling procedure there is only a coupling between bath 1 and the QD during step 1, hence $I_\text{step 1}$ stays always positive. 
For short durations, both average currents show a linear behavior where $I_\text{pump}$ is limited by the weak coupling between the QD and the second bath (cf. Section \ref{sec:projecMea}), while $I_\text{step 1}$ is limited by the strong coupling between the QD and the first bath and is about $\frac{\Gamma_1 \Delta_1}{\Gamma_2 \Delta_2}$ times larger%
\footnote{Immediately after a measurement (Zeno regime) or at the begin of step 1, the current between the QD and the reaction coordinate of bath $\nu$ increases linearly with time $t$ and behaves like $\sim \lambda_\nu^2 t $ with $\lambda_\nu=\sqrt{\Gamma_\nu \Delta_\nu/2}$, respectively. 
The ratio of the slopes corresponding to baths 1 and 2 is then $\frac{\lambda_1^2}{\lambda_2^2}=\frac{\Gamma_1 \Delta_1}{\Gamma_2 \Delta_2}$.} 
than $I_\text{pump}$.

\FloatBarrier
%%%%%%%%%%%%%%%%%%%%%%%%%%%%%%%%%%%%%%%%%%%%%%%%%%%%%%%%%%%%%%%%%%%%%%%%%%%%%%%%%%%%%%%%%%%%%%%%%%%%%%
\section{Conclusions}
In this work, we studied the influence of off--resonant tunneling and the related impact of the QD--bath coupling on the electron transport through quantum dots (QDs). We have focused on two electron pumps that pump electrons against an applied voltage and operate only in the higher--order tunneling regime. Both types of pumps consist of a single--level QD coupled to two fermionic baths at zero temperature and operate by decoherence operations and the corresponding resets of the coupling energy which drive off--resonant tunneling events.

In pumping scheme 1, we applied a coupling and decoupling procedure. It turned out that very structured baths may increase the efficiency with which the coupling/decoupling work is converted into the chemical energy gain of the second bath. Furthermore, finite cycle durations of the pumping procedure lead to strong non--Markovian effects, which correspond to oscillating electron occupations and coherences, and we used these to further increase efficiency. In addition, the presence of these oscillations is a clear signature for the importance of coherences between the QD and the bath  for the involved processes.

In pumping scheme 2, we considered a similar situation as before, but with periodically repeated projective measurements of the QD occupation and static couplings between the QD and the baths. For properly chosen bath structures and sufficiently frequent measurements, the measurements generate a current against the voltage applied between the baths.
In addition, we showed that even a weak continuous measurement by a charge detector, simulated by a single--electron transistor, can be sufficient to generate these currents. 
For both measurement realizations, the energy required for the pumping is gained by the measurements through the reset of the coupling energy and coherences between the QD and the baths for each measurement or during the continuous measurement and is similar to the anti--Zeno effect. Although the procedure is different, similar non--Markovian effects are observed as in the first pumping scheme and there is qualitative and quantitative agreement in the current flow between both schemes because both are driven by analogous decoherence operations and the manipulation of the coupling energy.    

%%%%%%%%%%%%%%%%%%%%%%%%%%%%%%%%%%%%%%%%%%%%%%%%%%%%%%%%%%%%%%%%%%%%%%%%%%%%%%%%%%%%
\FloatBarrier

\section*{Acknowledgments}
The authors thank Johann Zöllner for fruitful discussions and valuable feedback on the manuscript.
We gratefully acknowledge funding by the Deutsche Forschungs\-gemeinschaft (DFG, German Research Foundation) --- Project 278162697 --- SFB 1242.

\section*{Data Availability}
All data used and plotted in the figures are freely available in Ref. \cite{DUEDATA-2026-18LQAK_2026} (\href{https://doi.org/10.71955/DUEDATA-2026-18LQAK}{https://doi.org/10.71955/DUEDATA-2026-18LQAK}).
%%%%%%%%%%%%%%%%%%%%%%%%%%%%%%%%%%%%%%%%%%%%%%%%%%%%%%%%%%%%%%%%%%%%%%%%%%%%%%%%%%%%
\appendix

%%%%%%%%%%%%%%%%%%%%%%%%%%%%%%%%%%%%%%%%%%%%%%%%%%%%%%%%%%%%%%%%%%%%%%%%%%%%%%%%%%%%
\section{Exact solution for time--dependent systems} \label{sec: AppendixAexact}
 
\subsection{General derivation} \label{sec: nonhermitianOperatorAppendix}

In this section, we will extend the Heisenberg equation and the Laplace transform technique \cite{Topp_2015} to time--dependent Hamiltonians and we will reach an exact solution that has previously been derived by other methods \cite{jauho1994time,gurvitz_single_electron_2015}. 
With Eq. \eqref{exactsolution} and Eq. \eqref{M(t)} below it is possible to calculate the time dependence of all relevant expectation values.  

We start with considering a total Hamiltonian
\begin{multline}
    H(t)=\widetilde{H}_\text{S}(t)+ \sum_{k,\nu}\varepsilon_{\nu,k}a^\dagger_{\nu,k}a_{\nu,k} \\
    +\sum_{k,\nu}(t_{\nu,k} a^\dagger_{\nu,k}\, c_{\nu}+\text{h.c.})
    \label{H(t)}
\end{multline}
with time--independent tunneling matrix elements $t_{\nu,k}$ that satisfy the wideband limit. 
We note that, in this general formulation, each level $\nu$ of the system (with fermionic annihilation operator $c_\nu$) may be coupled to its own bath (with fermionic annihilation operators $a_{\nu,k}$).
In contrast to \cite{Topp_2015}, we allow the Hermitian system Hamiltonian $\widetilde{H}_\text{S}(t)$ to be time dependent.
The Heisenberg equations of motion for the annihilation operators are given by
 \begin{multline} \label{partialtcm}
      \partial_t c_{\nu,\text{H}}(t)=i [\widetilde{H}_{\text{S,H}}(t),c_{\nu,\text{H}}(t)]
    -  i\sum_k t^*_{\nu, k} a_{\nu,k,\text{H}}(t)
 \end{multline}
and
 \begin{equation} 
     \partial_t a_{\nu,k,\text{H}}(t)=-i\varepsilon_{\nu,k} a_{\nu,k,\text{H}}(t)-it_{\nu,k}c_{\nu,\text{H}}(t)
     \label{partialtamk}
 \end{equation}
with $\partial_t \equiv \frac{\mathrm{d}}{\mathrm{d}t}$.
 The index H indicates that the operator is in the Heisenberg picture $A_\text{H}(t)=U^\dagger(t)A(t)U(t)$ with the time evolution operator $U(t)$ obeying the differential equation
$\partial_tU(t)=-i H(t) U(t)$.
The Laplace transformation, $\mathcal{L}[g(t)](z)=\int_0^\infty\text{d}t \text{ }g(t)e^{-zt}$, of Eqs. \eqref{partialtcm}, \eqref{partialtamk}
gives
\begin{multline}\label{Lc(z)}
    \mathcal{L}[\partial_t c_{\nu,\text{H}}(t)](z)=i \mathcal{L}\Bigl[ [\widetilde{H}_{\text{S,H}}(t),c_{\nu,\text{H}}(t)]\Bigr](z)\\
    -  i\sum_k t^*_{\nu,k} \mathcal{L}[a_{\nu,k,\text{H}}(t)](z)
\end{multline}
and
\begin{equation}\label{La(z)}
    \mathcal{L}[a_{\nu,k,\text{H}}(t)](z)=\frac{a_{\nu,k}-i t_{\nu,k}\mathcal{L}[c_{\nu,\text{H}}(t)](z)}{z+i\varepsilon_{\nu,k}}\,.
\end{equation}
In Eq.~\eqref{La(z)}, we have used the mathematical relation $\mathcal{L}[\partial_t a_{\nu,k,\text{H}}(t)](z) = z \mathcal{L}[a_{\nu,k,\text{H}}(t)](z) - a_{\nu,k,\text{H}}(0)$ and the fact that $a_{\nu,k,\text{H}}(0) = a_{\nu,k}$
refers back to the Schr\"odinger picture.

Next, we insert Eq.~(\ref{La(z)}) into Eq.~(\ref{Lc(z)}).
Thereby, we encounter the sum
\begin{equation}
    A(z):= \sum_k \frac{|t_{\nu,k}|^2}{z+i\varepsilon_{\nu,k}}=\int_{-\infty}^{\infty}\frac{\text{d}\omega}{2\pi}\frac{\Gamma_{\nu}(\omega)}{z+i\omega}
    \label{A(z)}
\end{equation}
with $\Gamma_\nu(\omega) = 2\pi \sum_k |t_{\nu,k}|^2 \delta(\varepsilon_{\nu,k} - \omega)$.
In the wideband limit, i.e., for an energy--independent spectral--coupling density $\Gamma_\nu(\omega)\equiv \Gamma_\nu$, Eq.~(\ref{A(z)}) simplifies to the $z$--independent constant $A=\Gamma_\nu/2$.
This yields
\begin{multline}\label{Laplacec(z)}
    \mathcal{L}[\partial_t c_{\nu,\text{H}}(t)](z)=i \mathcal{L}\Bigl[ [\widetilde{H}_{\text{S,H}}(t),c_{\nu,\text{H}}(t)]\Bigr](z)\\
    -\frac{\Gamma_\nu}{2}\mathcal{L}[c_{\nu,\text{H}}(t)](z)-i\sum_k \frac{t_{\nu,k}^* a_{\nu,k}}{z+i\varepsilon_{\nu,k}}  \, .
\end{multline}
Since (as a consequence of the wideband limit) $z$ appears explicitly only in the last term, we can easily perform the inverse Laplace transformation.
We arrive at
\begin{equation}
    \partial_t c_{\nu,\text{H}}(t)=i[H_\text{eff,H}(t),c_{\nu,\text{H}}(t)]- i\sum_k t_{\nu,k}^* e^{-i\varepsilon_{\nu,k}t}a_{\nu,k}    
    \label{effkom}
\end{equation}
with an effective non--Hermitian operator 
\begin{equation}
    H_\text{eff}(t)=\widetilde{H}_\text{S}(t)-i\sum_{\nu}  \frac{\Gamma_\nu}{2} c_\nu^\dagger c_\nu
\end{equation}
in the Schr\"odinger and
$H_\text{eff,H}(t)=U^\dagger(t)H_\text{eff}(t)U(t)$ in the Heisenberg picture.

If $\widetilde{H}_\text{S}(t)$ is bilinear in $c_\nu$ and $c_\nu^\dagger$ with $\nu=1,\ldots, N$, Eq.~(\ref{effkom}) can be explicitly written as 
\begin{multline}
       \partial_t
            \begin{pmatrix}
            c_{1, \text{H}}(t)\\
            c_{2, \text{H}}(t)\\
            \vdots\\
            c_{N, \text{H}}(t)
            \end{pmatrix}
            =\textbf{M}(t)
            \begin{pmatrix}
            c_{1, \text{H}}(t)\\
            c_{2, \text{H}}(t)\\
            \vdots\\
            c_{N, \text{H}}(t)
            \end{pmatrix}
            \\- i\sum_k 
            \begin{pmatrix}
            t_{1,k}^* e^{-i\varepsilon_{1,k}t}a_{1,k}\\
            t_{2,k}^* e^{-i\varepsilon_{2,k}t}a_{2,k}\\
            \vdots\\
            t_{N,k}^* e^{-i\varepsilon_{N,k}t}a_{N,k}
            \end{pmatrix}
\end{multline}
where the time--dependent $N\times N$ matrix $\textbf{M}(t)$ is derived from the commutators in Eq.~(\ref{effkom}).
The solution of the differential equation is
\begin{multline}
    \begin{pmatrix}
            c_{1, \text{H}}(t)\\
            c_{2, \text{H}}(t)\\
            \vdots\\
            c_{N, \text{H}}(t)
    \end{pmatrix}
    = \mathbf{O}(t)    
    \begin{pmatrix}
            c_{1}\\
            c_{2}\\
            \vdots\\
            c_{N}
    \end{pmatrix}
    \\
    -i \mathbf{O}(t) \int_0^t\text{d}s\sum_k \mathbf{O}^{-1}(s)
            \begin{pmatrix}
            t_{1,k}^* e^{-i\varepsilon_{1,k}s}a_{1,k}\\
            t_{2,k}^* e^{-i\varepsilon_{2,k}s}a_{2,k}\\
            \vdots\\
            t_{N,k}^* e^{-i\varepsilon_{N,k}s}a_{N,k}
            \end{pmatrix}
            \label{exactsolution}
\end{multline}
with the propagator $\mathbf{O}(t)$ obeying $\partial_t\mathbf{O}(t)=\textbf{M}(t)\mathbf{O}(t)$, with $\mathbf{O}(0)=1$. 

This exact solution is only applicable to situations with time--independent $t_{\nu,k}$ in the wideband limit ($\Delta \rightarrow \infty$).
In order to use this technique for our studies as well, we employ the fermionic reaction--coordinate mapping as described in Section~\ref{sec:RC}.
For this, $\textbf{M}(t)$ is given by the $3 \times 3$ matrix
\begin{equation}
\label{M(t)}
\begin{split}
   \textbf{M}(t)= -&
   \begin{pmatrix}
    i\varepsilon& 0&0\\
       0&\Delta_1 +i\omega_1& 0\\
     0&0& \Delta_2+i\omega_2\\
   \end{pmatrix}\\
   &-  g_1(t)\sqrt{\frac{\Gamma_1 \Delta_1}{2}}
   \begin{pmatrix}
    0&i&0\\
       i&0& 0\\
    0&0& 0\\
   \end{pmatrix}
   \\&-  g_2(t)\sqrt{\frac{\Gamma_2 \Delta_2}{2}}
   \begin{pmatrix}
    0&0&i\\
       0&0& 0\\
    i&0& 0\\
   \end{pmatrix} \, ,
    \end{split}
\end{equation}
where the first row/column refer to the QD level ($c_1 \equiv c$), the second to reaction--coordinate 1 ($c_2 \equiv r_1$) and the third to reaction--coordinate 2 ($c_3 \equiv r_2$).
The fact that, after the mapping, the QD orbital is no longer coupled to the bath is accounted for by setting $t_{1,k}=0$.
The coupling of reaction--coordinate 1 and 2 to their respective bath, $t_{2,k} \equiv \tilde{\gamma}_{1,k}$ and $t_{3,k} \equiv \tilde{\gamma}_{2,k}$ with $a_{2,k}=\tilde{a}_{1,k}$ and $a_{3,k}=\tilde{a}_{2,k}$, leads to the (transformed) spectral--coupling density $\tilde{\Gamma}_\nu (\omega)= 2 \pi \sum_k |\tilde{\gamma}_{\nu,k}|^2 \delta (\omega - \tilde{\varepsilon}_{\nu,k})=2 \Delta_\nu$, which satisfies the wideband condition.
The imaginary parts of $\textbf{M}(t)$ represent the energy levels of $\widetilde{H}_\text{S}(t)$, while the real parts of $\textbf{M}(t)$ correspond to the decay rates and the effective broadening of the energy levels due to higher--orders in the tunnel coupling.

With Eq.~\eqref{exactsolution} and Eq.~\eqref{M(t)}, we know the creation/annihilation operators in the Heisenberg picture as a function of time and the operators in the Schrödinger picture, and we are able to calculate the time dependence of the expectation values of all observables for given initial conditions. 
    
%%%%%%%%%%%%%%%%%%%%%%%%%%%%

\subsection{Stationary QD occupation}%
\label{sec: exactAppendixstationaryQDn}%
For long equilibration times $t_1,\,t_2 \rightarrow \infty$, the QD will equilibrate with the bath to which it is coupled (bath $\nu=1 \rightarrow$ step 1; bath $\nu=2 \rightarrow$ step 2). 
In this regime, from the structure of the matrix $\textbf{M}(t)$, it follows that the first term of Eq.~\eqref{exactsolution} vanishes. 
Therefore, the initial values for the QD and reaction--coordinates operators in Eq. \eqref{exactsolution} can be omitted. With Eq. \eqref{M(t)} and initial bath--mode occupations that are given by the Fermi function $f_+(\omega,\mu_\nu)=1/[e^{\beta (\omega-\mu_\nu)}+1]$, for the inverse temperature $\beta$, chemical potentials $\mu_\nu$ and energy $\omega$, we obtain the stationary QD occupation
\begin{equation}
    \left<n_{\text{stat},\nu}\right>=  \int \limits_{-\infty}^\infty \frac{\text{d}\omega}{2 \pi} \frac{ \Gamma_\nu f_+(\omega,\mu_\nu)}{(\varepsilon-\omega)^2+\left(\frac{\Gamma_\nu}{2}+\frac{(\varepsilon-\omega)(\omega-\omega_\nu)}{\Delta_\nu}\right)^2}\,.
\end{equation}
In the limit of weak coupling, $\Gamma_1\searrow0$, the stationary QD occupation is given by the Fermi function $f_+(\varepsilon,\mu_\nu)$ at the energy level of the QD.
If $\Gamma_1$ is sufficiently large compared to $\Gamma_2$, it is possible to end up with $\left<n_{\text{stat},1}\right> > \left<n_{\text{stat},2}\right>$ even though $\mu_2 > \mu_1$.

%%%%%%%%%%%%%%%%%%%%%%%%%%%%

%%%%%%%%%%%%%%%%%%%%%%%%%%%%%%%%%%%%%%%%%%

\section{Energy efficiency of pumping scheme 1}
In this appendix, we discuss the energy efficiency of pumping scheme 1 in the regime of extremely weak coupling $\Gamma_2$ between the QD and the second bath ($\Gamma_2 \ll \varepsilon -V\ll V$ and $\Gamma_2\ll \Gamma_1$) and long durations of the second step (formally $t_2 \rightarrow \infty$). 
In this regime, the work $W_\text{C}(t_1)$ (cf. Eq. \eqref{WC(t)}) performed by the coupling/decoupling procedure changes the energy $\Delta E_\text{QD} = \varepsilon\, n_\text{QD}(t_1)$ of the QD and bath 1 ($\Delta E_\text{B}$), while the bath 2 remains unaffected. 
Then, the energy efficiency (cf. Eq. \eqref{etapump})
\begin{equation}\label{etaAppB1}
   \eta_\text{pump} \approx
   \frac{\varepsilon\, n_\text{QD}(t_1)}{W_\text{C}(t_1)} = 
   \frac{\Delta E_\text{QD}}{\Delta E_\text{B} + \Delta E_\text{QD}}
\end{equation}
quantifies 
the energy gain of the QD ($\Delta E_\text{QD}$), used for pumping, as a part of the total work $W_\text{C}(t_1) = -\left<H_\text{C}(t_1-0^+)\right> =\Delta E_\text{B} + \Delta E_\text{QD}$, 
while the energy excitations of the bath ($\Delta E_\text{B}$) do not contribute to pumping.
The energy efficiency is, therefore, traced back to how the performed work is distributed between QD and bath. 
In Subsection \ref{sec:appExcitaionsBath}, we use an exact solution to discuss how exactly the performed work is distributed between different areas of the bath and how this influences the energetic efficiency. The details of the exact solution are provided in Subsection \ref{sec:appCalculationBath}, while in Subsection \ref{sec: exactsoMstep1_eta}, we derive a simple estimate for the energetic efficiency from the ground state of the supersystem after the reaction--coordinate mapping for $t_1,\,t_2\rightarrow\infty$.

\subsection{Relation between energetic efficiency and bath excitations}\label{sec:appExcitaionsBath}

To resolve the bath degrees of freedom further, we introduce the energy gain density 
\begin{equation}\label{EB(w)}
    \Delta E_\text{B}(\omega)=\sum_k \varepsilon_{1,k}\Bigl[\left<n_{1,k}(t_1)\right>-\left<n_{1,k}(0)\right>\Bigr] \delta(\varepsilon_{1,k}-\omega)
\end{equation}
that describes the gain in mode $\omega$ of the bath
and consider separately the part
of the bath states below the Fermi level
\begin{equation}\label{EB<}
\Delta E_{\text{B}, <0}= \int_{-\infty}^0 \text{d} \omega \, \Delta E_\text{B}(\omega)
\end{equation}
and above the Fermi level
\begin{equation}\label{EB>}
\Delta E_{\text{B}, >0}= \int_{0}^\infty \text{d} \omega \, \Delta E_\text{B}(\omega)\, ,
\end{equation}
which add up to the total energy gain of bath,
$\Delta E_\text{B}=\Delta E_{\text{B}, <0}+\Delta E_{\text{B}, >0}$.

\begin{figure}
    \centering
    \includegraphics[width=\linewidth]{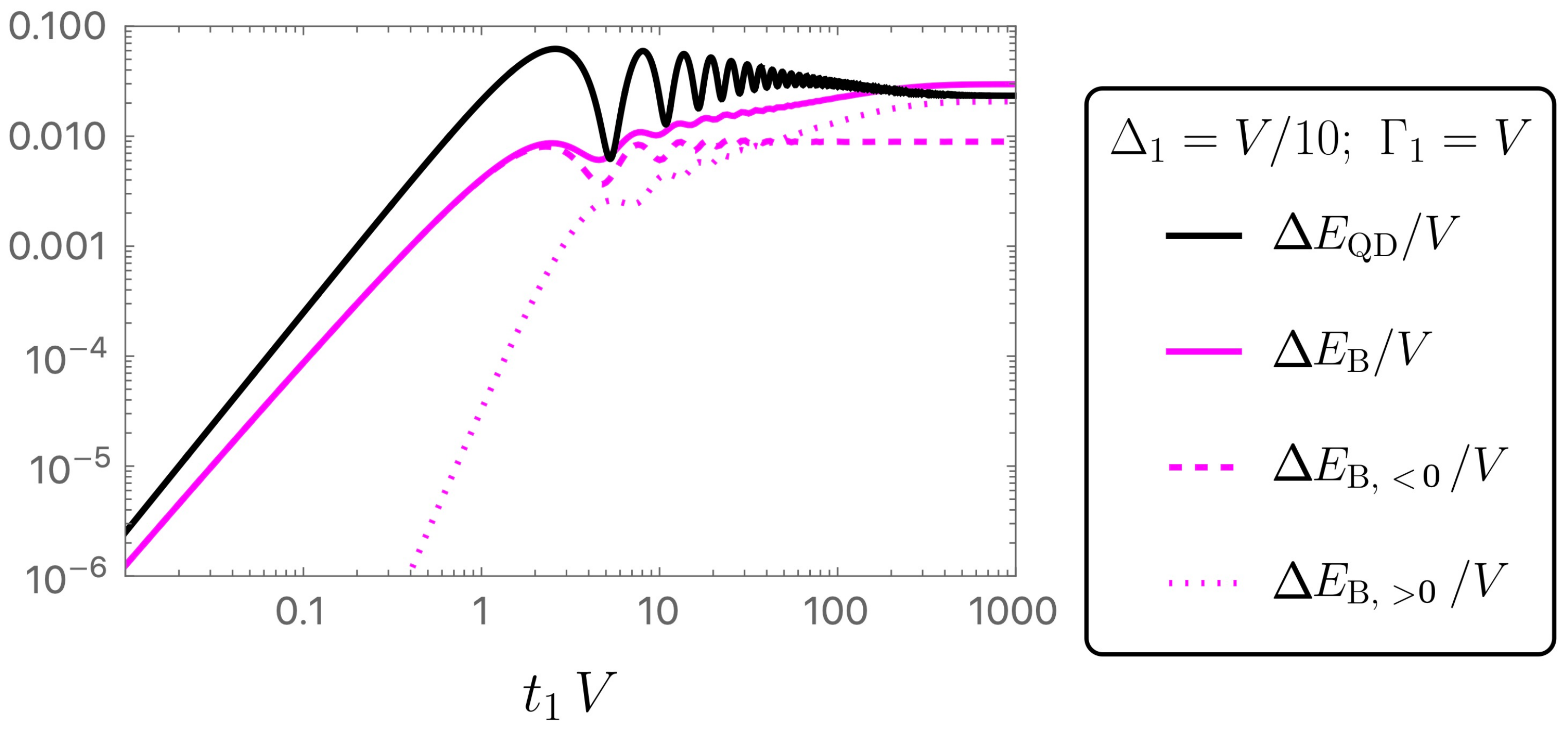}
    \caption{The energy gain $\Delta E_\text{QD}$ of the QD and of the bath $\Delta E_\text{B}=\Delta E_{\text{B}, <0}+\Delta E_{\text{B}, >0}$ during step 1 are shown as functions of the duration $t_1$ of step 1. The energy gain of the bath is the sum of the energy gain $\Delta E_{\text{B}, <0}$ induced by electrons that leave the bath from states below the Fermi level and the energy gain $ \Delta E_{\text{B}, >0}$ induced by electrons that enter the bath to states above the Fermi level. For short $t_1$, $\Delta E_\text{QD}$ behaves like $\sim t_1^2$ and $\Delta E_{\text{B}, <0}$ behaves approximately like $\sim t_1^2 \log t_1$. In contrast to this, $\Delta E_{\text{B}, >0}$ behaves like $\sim t_1^4 \log t_1$. The parameters are $\Delta_1=V/10$, $\Gamma_1=V$, $\omega_1=0$, $\Gamma_2 \ll \varepsilon-V\ll V$ (formally $\varepsilon \searrow V$), $\Gamma_2\ll \Gamma_1$ and $t_2\rightarrow \infty$. 
    }
    \label{fig:DEbath_DVd10uGV}
\end{figure}

\begin{figure}
    \centering
    \includegraphics[width=\linewidth]{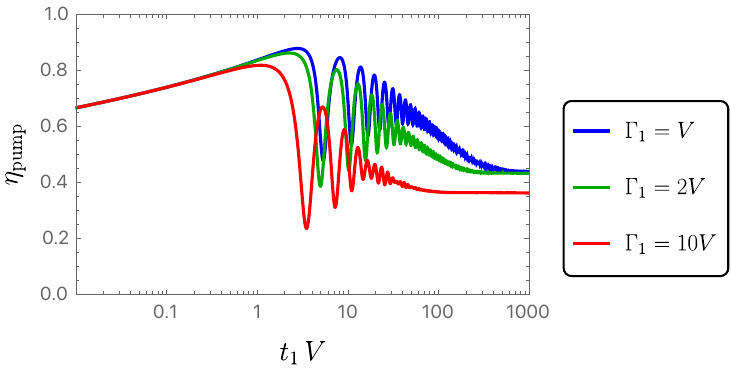}
        \includegraphics[width=\linewidth]{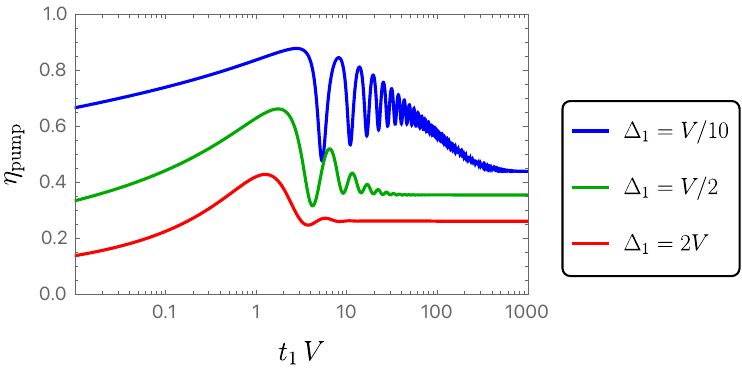}
    \caption{The energy efficiency $\eta_\text{pump}$ (top and bottom) is shown as a function of the duration $t_1$ of step 1. The parameters are $\omega_1=0$, $\Gamma_2 \ll \varepsilon-V\ll V$ (formally $\varepsilon \searrow V$), $\Gamma_2\ll \Gamma_1$ and $t_2\rightarrow \infty$.
    Top: The different curves correspond to different $\Gamma_1$ but identical $\Delta_1=V/10$.
    Bottom: The different curves correspond to different $\Delta_1$ but identical $\Gamma_1=V$. 
    }
    \label{fig:eta_bath_DVd10uGlauf_t1}
\end{figure}

Figure \ref{fig:DEbath_DVd10uGV} shows the different contributions $\Delta E_\text{QD}$, $\Delta E_{\text{B}, <0}$, $\Delta E_{\text{B}, >0}$, and $\Delta E_\text{B}$ as functions of the duration $t_1$.
Technical details of the calculations are provided in the next subsection.
For short $t_1$, the number of electrons leaving the bath, here $n_\text{QD}(t_1)=\Delta E_\text{QD}/\varepsilon$, increases quadratically with $t_1$, 
\begin{equation}
    n_\text{QD}(t_1)\sim t_1^2 \qquad \text{(for $t_1\sqrt{\Gamma_1 \Delta_1/2}\ll 1$)}.
\end{equation}
In combination with 
\begin{equation}\label{EB<smallt1}
    \Delta E_{\text{B}, <0}\approx -\Bar{\varepsilon}(t_1)n_\text{QD}(t_1)
\end{equation}
and the averaged energy of the leaving electrons, $\Bar{\varepsilon}(t_1)\sim \log t_1<0$, it follows that $\Delta E_{\text{B}, <0} \sim -t_1^2 \log t_1$. Excitations to states above the Fermi level are only possible by including the QD as an intermediate step, and the corresponding energy $\Delta E_{\text{B}, >0}\sim -t_1^4 \log t_1$ increases slower than $\Delta E_{\text{B}, <0}$ for short $t_1$. Therefore, $\Delta E_{\text{B}, <0}$ dominates in this regime and the efficiency (cf. Eq. \eqref{etaAppB1}) is given by 
\begin{equation}\label{etasmallt1}
    \eta_\text{pump}\approx \frac{\varepsilon}{\varepsilon-\Bar{\varepsilon}(t_1)} 
\end{equation}
where $\Bar{\varepsilon}(t_1)$ depends only on $\Delta_1$ and $t_1$ and, hence, $\eta_\text{pump}$ becomes independent of $\Gamma_1$ (cf. Figure \ref{fig:eta_bath_DVd10uGlauf_t1} --- top). For long $t_1$, electrons entering the QD above the Fermi level (corresponding to $\Delta E_{\text{B}, >0}$) become relevant and lead to a reduction of $\eta_\text{eff}$ compared to short $t_1$ where the reduction increases with increasing $\Gamma_1$ or $\Delta_1$ (Figure \ref{fig:eta_bath_DVd10uGlauf_t1}). 

\begin{figure}
    \centering
    \includegraphics[width=\linewidth]{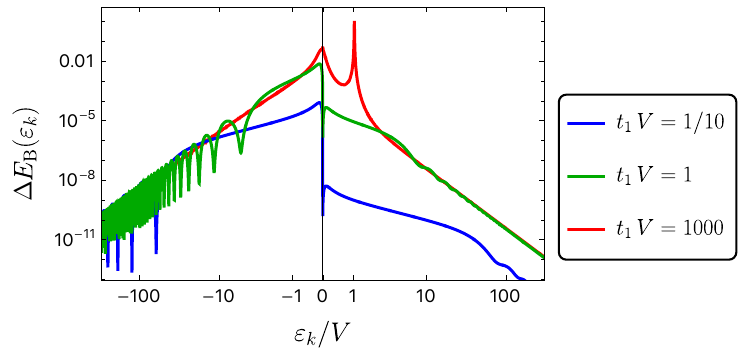}
    \caption{The energy density $\Delta E_\text{B}(\varepsilon_k)$ (cf. Eq. \eqref{EB(w)}) corresponding to the change in bath energy during step 1 is shown as a function of $\varepsilon_k$ for three different values of $t_1$. The vertical axis is scaled logarithmically and the horizontal axis is scaled with $\text{arsinh}(\cdot)$, resulting in a double--log plot for $|\varepsilon_k|/V\gg1$. The parameters are the same as in Figure \ref{fig:DEbath_DVd10uGV}.}
    \label{fig:Ebath(ek)_DVd10uGV}
\end{figure}

Figure \ref{fig:Ebath(ek)_DVd10uGV} shows $ \Delta E_\text{B}(\varepsilon_k)$ as a function of the energy $\varepsilon_k$ of the corresponding bath mode for different durations $t_1$. For long $t_1$, $\Delta E_\text{B}(\varepsilon_k)$ is localized at the Fermi level and at the energy level of the QD. The peak, corresponding to the energy level of the QD, vanishes for smaller $t_1$ when $\Delta E_{\text{B}, >0}$ is not dominant.  The flanks of $\Delta E_\text{B}(\varepsilon_k)$ decay as $\sim |\varepsilon_k|^{-3}$. In a range of approximately $|\varepsilon_k|\lesssim \frac{1}{t_1}$, the flanks show a suppressed decay $\sim |\varepsilon_k|^{-1}$ for short $t_1$. This flattening leads to the time dependence of $\bar{\varepsilon}(t_1)$ and $\eta_\text{pump}$ for short $t_1$ (cf. Eq. \eqref{EB<smallt1} and Eq. \eqref{etasmallt1}).

\subsection{Calculation details}\label{sec:appCalculationBath}
For $\Gamma_2 \ll \varepsilon -V\ll V$, $\Gamma_2\ll \Gamma_1$ and $t_2 \rightarrow \infty$, the density matrix $\rho_0$ of the total system, i.e., QD~+~baths, at the beginning of step 1 is a product state between the ground state%
\footnote{This means $\tr \left( \rho_0 a_{1,k}^\dagger a_{1,q}\right)=\delta_{k,q}\Theta(-\varepsilon_{1,k})$ and this is not identical to $\tr \left( \rho_0 \tilde{a}_{1,k}^\dagger \tilde{a}_{1,q}\right)=\delta_{k,q}\Theta(-\tilde{\varepsilon}_{1,k})$ (cf. Section~\ref{sec:RC}).}
of bath 1
and the state of an empty QD. 
In this case, it would be an unnecessary complication to use Eq. \eqref{exactsolution} together with Eq. \eqref{M(t)} after a reaction--coordinate mapping, since in terms of the transformed degrees of freedom the initial state is no longer a product state.
Instead, we repeat the procedure for deriving the exact solution presented in Appendix \ref{sec: nonhermitianOperatorAppendix}, but this time assuming a time--independent system Hamiltonian (since the tunnel couplings are kept constant during the steps), while taking into account the Lorentzian--shaped spectral coupling density for the coupling between QD and bath 1.
Up to Eqs.~\eqref{Lc(z)} and \eqref{La(z)}, the derivation remains the same with only one system degree of freedom, $\nu=1$.
For Eq.~\eqref{Lc(z)}, we use the relation $\mathcal{L}[\partial_t c_{1,\text{H}}(t)](z) = z \mathcal{L}[c_{1,\text{H}}(t)](z) - c_1$ and explicitly calculate the commutator $[\widetilde{H}_{\text{S,H}}(t),c_{1,\text{H}}(t)]= - \varepsilon c_{1,\text{H}}(t)$.
The sum in Eq.~\eqref{A(z)}, that appears when inserting Eq.~(\ref{La(z)}) into Eq.~(\ref{Lc(z)}), preserves its $z$ dependence, 
\begin{equation}
    A(z)=\frac{1}{2} \frac{\Gamma_1 \Delta_1}{z+i\omega_1+\Delta_1 } \, ,
\end{equation}
and is, in contrast to the wideband limit, no longer constant.
As a result, we have to replace Eq.~\eqref{Laplacec(z)} by
\begin{equation}\label{L(z)_Lorentz}
    \mathcal{L}[ c_{1,\text{H}}(t)](z)=\frac{1}{z+i\varepsilon +\frac{1}{2}\frac{\Gamma_1 \Delta_1}{z+i\omega_1 +\Delta_1}}\left(c_1-i\sum_k\frac{\gamma_{1,k}^* a_{1,k}}{z+i \varepsilon_{1,k}}\right) \, .
\end{equation}
It is now straightforward to perform the inverse Laplace transformation on Eq.~\eqref{L(z)_Lorentz} and Eq.~\eqref{La(z)} to get $a_{1,k,\text{H}}(t)$ and $c_{1,\text{H}}(t)$ as functions of $t$, $c_1$ and $a_{1,k}$.
With this, we are able to calculate the time dependence of the particle number of the QD $\left<n_\text{QD}(t)\right>=\tr \left( \rho_0 c_{1,H}^\dagger(t) c_{1,H}(t)\right)$ and of the bath states $\left<n_{1,k}(t)\right>=\tr \left( \rho_0 a_{1,k,H}^\dagger(t) a_{1,k,H}(t)\right)$ by using $\tr \left( \rho_0 c_{1}^\dagger c_{1}\right)=\tr \left( \rho_0 c_{1}^\dagger a_{1,k}\right)=0$ and $\tr \left( \rho_0 a_{1,k}^\dagger a_{1,q}\right)=\delta_{k,q}\Theta(-\varepsilon_{1,k})$. Here, the occupation $\left<n_{1,k}(t)\right>$ of each mode $k$ changes only by a small value in time and vanishes in the continuum limit. However, the sum over all bath states leads to a finite change in the number of particles in the bath. 

%%%%%%%%%%%%%%%%%%%%%%%%%%%%%%%%%%%%%

\subsection{Simple estimate for slow pumping}
\label{sec: exactsoMstep1_eta}
In the following we consider the slow pumping regime ($t_1,\,t_2\rightarrow\infty$) of pumping scheme 1 discussed in Section \ref{sec:slowpumping} for $\Gamma_2 \ll \varepsilon -V$ and $\Gamma_2\ll \Gamma_1$. For $\omega_1\leq \frac{\Gamma_1 \Delta_1}{2}$, the number of pumped electrons $N_\text{pump}$ and the coupling/decoupling work can be calculated by approximating the stationary states with the ground states of $\widetilde{H}_\text{S}(t)$ (cf. Eq.~\eqref{tilde_H_S}).
For $\omega_1< \frac{\Gamma_1 \Delta_1}{2}$, the ground state during step 1 is given by a superposition of the state $\ket{\mathrm{RC1}}$ where only the first reaction coordinate is occupied and the state $\ket{\mathrm{QD}}$ where only the QD is occupied, 
\begin{equation}
    \left|E^\mathrm{QD-RC1}_-\right>=\frac{\kappa_- \ket{\mathrm{RC1}}+\ket{\mathrm{QD}}}{\sqrt{1+\kappa_-^2}}
\end{equation}
with $\kappa_-=(\omega_1-\varepsilon-\sqrt{(\varepsilon-\omega_1)^2+ 2 \Delta_1 \Gamma_1})/({\sqrt{2 \Delta_1 \Gamma_1}})$. In contrast, the ground state during step 2 is the unoccupied QD. Based on this, we can calculate the energetic efficiency
\begin{equation}
    \eta_\text{pump}\approx \frac{V}{\varepsilon-\omega_1+\sqrt{(\varepsilon-\omega_1)^2+2 \Gamma_1 \Delta_1}}
\end{equation}
for $0\leq\omega_1\leq \frac{\Gamma_1 \Delta_1}{2}$ (cf. Eq. \eqref{etaAppB1}). 
For $\varepsilon \searrow V$ and $\omega_1 \nearrow \frac{\Gamma_1 \Delta_1}{2}$ or $\Delta_1 \Gamma_1 \rightarrow 0$ for $\omega_1=0$, the efficiency reaches the maximum value $\eta_\text{pump}=\frac{1}{2}$.

\section{Stroboscopic projective measurement}

\subsection{Implementation}
\label{sec:exactwithmeasurement}

A projective measurement of the particle number of the QD with measurement outcome $n$ is given by the superoperator $\mathcal{P}_n$ that maps the density matrix $\rho$ onto
\begin{equation}\label{Pmeasure}
    \mathcal{P}_n[\rho]=\frac{P_n\rho P_n}{p_n} \, .
\end{equation}
Here, $P_0=c c^\dagger$ is the projector for the measurement outcome ``$0$'', $P_1=c^\dagger c$ the projector for the measurement outcome ``$1$'', and
\begin{equation}
    p_n=\text{tr}\left(P_n \rho\right)
\end{equation}
is the probability to measure the QD occupation $n$.
Averaging over all measurement outcomes leads to the density matrix
\begin{equation}
    \overline{\rho}=\sum_n p_n \mathcal{P}_n[\rho]
    = P_{0}\rho P_{0}+P_{1}\rho P_{1} \, .
\end{equation}
In consequence, each measurement deletes coherences between the QD and the baths. 
However, the  coherences between the baths and the  coherences inside the baths remain unaffected.
This point will be illustrated in Appendix \ref{sec: appendixmeasurementwideband}. 

The immediate destruction of the coherence between QD and bath due to a projective measurement is implemented into the exact--solution scheme described in Appendix~\ref{sec: nonhermitianOperatorAppendix} with the following trick.
We formally introduce two copies of the QD orbital, described by the annihilation operators $c_1$ and $c_2$.
Before the measurement, only orbital 1 is coupled to the baths, while orbital 2 is uncoupled.
The measurement process is modeled by decoupling orbital 1 and, at the same time, coupling orbital 2 to the (same) baths and using the initial condition that orbital 2 has the same occupation as orbital 1, $\left<c^\dagger_2 c_2\right>=\left< c^\dagger_1 c_1\right>$, at the time of measurement.
In that way, the information about the QD occupation is transferred from the first to the second copy of the orbital, but the coherences between QD and baths are lost.
In contrast, coherences within the baths remain.
The fact, that such coherences can have measurable effects is demonstrated in the next subsection.

Now we know how to implement one projective measurement into the exact--solution scheme. For more than one measurement, the discussed trick can be repeated with the number of required QD copies given by the number of measurements.

%%%%%%%%%%%%%%%%%%%%%%%%%%%%%%%%%%%%%%%%%%%%%

\subsection{Effect of intra--bath coherences}
\label{sec: appendixmeasurementwideband}

To illustrate the relevance of coherences within the baths (intra--bath coherences), we consider in this subsection a simple case scenario where a QD is in equilibrium with one bath with a wideband spectral coupling density, $\Delta \to \infty$.
The equilibrium condition implies that, on average, there is no current flowing between QD and bath.
The average QD occupation is some number between 0 and 1.
After a projective measurement, the QD occupation is either $n=0$ or $n=1$.
In the subsequent time evolution, the QD occupation reaches again its average value, i.e., current is flowing transiently between QD and bath.
The detailed time dependence of this current depends on whether intra--bath coherences are taken into account or not.

\begin{figure}
    \centering
    \includegraphics[width=\linewidth]{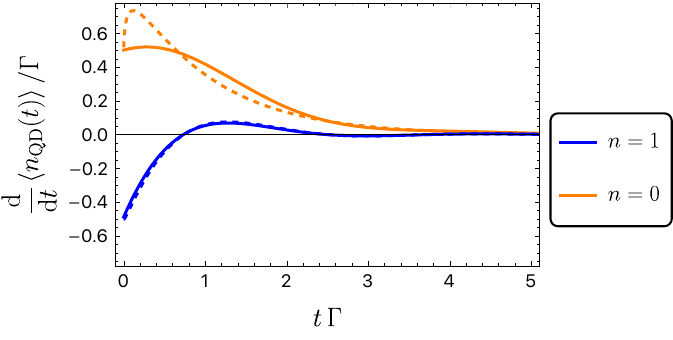}
    \caption{The total current is shown as a function of time. For the dashed curves, the initial state is prepared by measuring the occupation of the QD. In contrast, the solid {curves} indicate the situation where the initial state of the bath is given by a state where no  coherences between different bath modes are present. For an initially unoccupied QD (orange color), there are significant differences between the total currents of the two different bath states, whereas for an initially occupied QD (blue color), the total currents
    for the two different prepared initial states
    are similar. The parameters are $\varepsilon=\mu - 2\Gamma$ are $\Delta \rightarrow \infty$ (wideband).}
    \label{fig:withMeasurementn}
\end{figure}

This is illustrated in Figure \ref{fig:withMeasurementn}, which shows the time derivative of the particle number of the QD (i.e., the current) as a function of time after having measured $n$ by a projective measurement (dashed curves).
This is compared to the case when bath is assumed to be in an equilibrium Gibbs state with no intra--bath coherences (solid curves).
We find that there are significant deviations between  the initially correlated and the initially uncorrelated bath.
For the chosen parameters, for which the QD is more likely to be occupied than empty, the deviations are more pronounced for $n=0$.

To understand on a formal level that the intra--bath coherences can matter, we calculate the time derivative of the current operator $\displaystyle\hat{I}_k=i \sum_{k}\left(\gamma_{k}a_{k}^\dagger c - \text{h.c.}\right)$ between the bath mode $k$ and the QD.
Making use of $\frac{\text{d} \hat{I}_k}{\text{d} t} = i [H ,\hat{I}_k]$, we obtain
\begin{align}
    \frac{\text{d} \hat{I}_k}{\text{d} t} 
    =& 2 |\gamma_k|^2\left( a^\dagger_k a_k -c^\dagger c\right)
    +(\varepsilon-\varepsilon_k)\left(\gamma_k  a_k^\dagger c + \mathrm{ h.c. }\right) \nonumber \\
    &+ \sum_{l\neq k}\left(\gamma_k^*\gamma_l a^\dagger_l a_k + \mathrm{ h.c. }\right) \, .
\end{align}
The first term, that depends on the occupation difference between QD and bath, is not affected by coherences.
The second term corresponds to coherences between QD and bath that are destroyed by the projective measurement.
The third term, however, corresponds to coherences between different bath modes.
As a consequence, these coherences can influence the time derivative of the current and, thus, the current itself.

%%%%%%%%%%%%%%%%%%%%%%%%%%%%%%%%%%%%%%%%%%

\section{Continuous weak measurement}
\label{sec:AppRedfield}

The setup for pumping scheme 2 with continuous weak measurement is described by the Hamiltonian Eq. \eqref{tildeHtot}.
After performing the reaction--coordinate mapping for the QD (but not the SET), we have to deal with the system, described by the Hamiltonian
\begin{multline}
   H'_\text{S}=Un \, n_\SET+\varepsilon n+\sum_\nu \omega_\nu r_\nu^\dagger r_\nu  \\
   + \sum_\nu  \sqrt{\frac{\Gamma_\nu \Delta_\nu}{2}}(r_\nu^\dagger \,c+\text{h.c.}) \, 
\end{multline}
with $n=c^\dagger c$ the occupation of the QD and the occupation $n_\SET$ of the SET.
The tunnel couplings of the reaction coordinates to the baths are described by the tunneling strengths $2\Delta_\nu$, and the SET orbital is coupled to its baths with strength $\Gamma_\SET$.
All these couplings are given in the wideband limit.

Due to the interaction term $U n\, n_\SET$, which is quartic in the fermionic operators, the exact solution following the procedure described in Appendix~\ref{sec: nonhermitianOperatorAppendix} is no longer available.
Therefore, we employ for this part, as an alternative, a systematic perturbation expansion in the tunnel couplings.
For this, we derive a master equation for the reduced density matrix $\rho_\text{S}$ of system S by integrating out the degrees of freedom of the baths and neglect terms beyond the first order in $\Gamma_\SET$ and $\Delta_\nu$.
The result is the 
Redfield--II equation
\cite{Redfield,B07-Trimer,Teff}
\begin{multline}\label{RedII}
    \partial_t \rho_\text{S} (t) = -i \left[H'_\text{S}+H'_\text{S,Lamb},\rho_\text{S} (t)\right]\\+ \sum_{\alpha,\nu}\sum_{\Delta E,\Delta E'}\frac{f_\alpha(\Delta E,\mu_\nu)+f_\alpha(\Delta E',\mu_\nu)}{2}\\
    \times
    \Biggl (K^\alpha_\nu(\Delta E)\, \rho_\text{S}(t)\, K^\alpha_\nu(\Delta E')^\dagger \\
	 - \frac{1}{2} \left\{ K^\alpha_\nu(\Delta E')^\dagger K^\alpha_\nu(\Delta E), \rho_\text{S}(t) \right\}\Biggr)
\end{multline}
with $\nu \in \{1,2,\text{source, drain}\}$, $\alpha \in \{\pm$\}, 
the renormalization (or Lamb--shift) term%
\footnote{Here, we neglect the imaginary prefactors in Eq. \eqref{RedII} and the real prefactors of the renormalized term in Eq. \eqref{Lampshift} for simplicity \cite{Teff}.  }
\begin{multline}\label{Lampshift}
    H'_\text{S,Lamb}=i \sum_{\substack{\nu,\alpha \\ \Delta E, \Delta E'}} \frac{f_{\alpha}(\Delta E',\mu_\nu) - f_{\alpha}(\Delta E,\mu_\nu)}{4}\\
    \times
	K^\alpha_{\nu}(\Delta E')^\dagger K^\alpha_{\nu}(\Delta E)
\end{multline}
and the jump operators
\begin{equation}
   K^\pm_{\nu}(\Delta E) = \sqrt{2\Delta_\nu}\sum_{k,l} \delta_{E_k-E_l,\pm \Delta E} \ketbra{E_k}{E_k} r_\nu^\pm \ketbra{E_l}{E_l}
\end{equation}
for $\nu=1,2$ and 
\begin{equation}
   K^\pm_{\nu}(\Delta E) = \sqrt{\Gamma_\SET}\sum_{k,l} \delta_{E_k-E_l,\pm \Delta E} \ketbra{E_k}{E_k} c_\SET^\pm \ketbra{E_l}{E_l}
\end{equation}
for $\nu=\text{source}, \text{ drain}$.
Here, $f_+(\Delta E,\mu_\nu)=1/[e^{\beta (\Delta E-\mu_\nu)}+1]$ is the Fermi function, in our case at zero temperature ($\beta\rightarrow \infty$), and $f_-(\Delta E,\mu_\nu)=1-f_+(\Delta E,\mu_\nu)$.
The states $\ket{E_l}$ are eigenstates of $H'_\text{S}$ with eigenenergies $E_l$.

We use Eq. \eqref{RedII} to find the stationary state $\rho_\text{\text{S,stat}}$ and calculate the pumped current
\begin{equation}
    I_\text{pump}=-2 \sqrt{\frac{\Gamma_1 \Delta_1}{2}} \mathrm{Im}\left<r_1^\dagger c\right>= 2 \sqrt{\frac{\Gamma_2 \Delta_2}{2}} \mathrm{Im}\left<r_2^\dagger c\right>
\end{equation}
and the current flowing through the SET
\begin{equation}
    I_\SET=\left<\mathcal{D}_\text{source} n_\SET\right>=-\left<\mathcal{D}_\text{drain} n_\SET\right>
\end{equation}
with
\begin{multline}
    \mathcal{D}_\nu n_\SET = \sum_{\alpha}\sum_{\Delta E,\Delta E'}\frac{f_\alpha(\Delta E,\mu_\nu)+f_\alpha(\Delta E',\mu_\nu)}{2}\\
    \times
    \Biggl (K^\alpha_\nu(\Delta E) ^\dagger\, n_\SET\, K^\alpha_\nu(\Delta E') \\
	 - \frac{1}{2} \left\{ K^\alpha_\nu(\Delta E')^\dagger K^\alpha_\nu(\Delta E), n_\SET \right\}\Biggr)\,.
\end{multline}
In Figure \ref{fig:SET_Ipump}, we vary $\Gamma_\SET$ and calculate $I_\SET$ and $I_\text{pump}$.
When calculating $\rho_\text{\text{S,stat}}$ it is important to take the renormalization term $H'_\text{S,Lamb}$ into account. Otherwise, there appear nonphysical currents between the SET and the QD, unrelated to any tunneling term in the Hamiltonian $H_\text{S}'$.  
Moreover, the Redfield equation is only an approximation for small $\Delta_\nu$ and small $\Gamma_\SET$ which leads to some errors, in particular,  $I_\text{pump}$ cannot become negative in the Redfield equation, while the exact solution enables negative currents.
It breaks down for large $\Delta_\nu$ or $\Gamma_\SET$.

\bibliography{Literatur}

\end{document}